\documentclass[12pt]{article}
\usepackage[dvips]{graphics,color}
\usepackage{latexsym,amssymb}
\usepackage{amsmath,amsbsy}
\usepackage[all]{xy}
\usepackage{amsopn,amstext}
\usepackage{cite}
\usepackage{amssymb}
\usepackage{rotating}
\usepackage{amsmath}
\usepackage{amsopn,amstext}

\allowdisplaybreaks[4]

\renewcommand{\theequation}{\arabic{section}.\arabic{equation}}

\begin{document}

%************************** Text Begins here ******************************

%  Greek letters

\def\a{\alpha}
\def\b{\beta}
\def\d{\delta}
\def\e{\epsilon}
\def\g{\gamma}
\def\h{\mathfrak{h}}
\def\k{\kappa}
\def\l{\lambda}
\def\o{\omega}
\def\p{\wp}
\def\r{\rho}
\def\t{\tau}
\def\s{\sigma}
\def\z{\zeta}
\def\x{\xi}
\def\V={{{\bf\rm{V}}}}
 \def\A{{\cal{A}}}
 \def\B{{\cal{B}}}
 \def\C{{\cal{C}}}
 \def\D{{\cal{D}}}
\def\G{\Gamma}
\def\K{{\cal{K}}}
\def\O{\Omega}
\def\R{\bar{R}}
\def\T{{\cal{T}}}
\def\L{\Lambda}

\def\R{\overline{R}}

\def\beq{\begin{equation}}
\def\eeq{\end{equation}}
\def\bea{\begin{eqnarray}}
\def\eea{\end{eqnarray}}
\def\ba{\begin{array}}
\def\ea{\end{array}}
\def\no{\nonumber}
\def\le{\langle}
\def\re{\rangle}
\def\lt{\left}
\def\rt{\right}

\baselineskip=20pt

\newcommand{\Title}[1]{{\baselineskip=26pt
   \begin{center} \Large \bf #1 \\ \ \\ \end{center}}}
\newcommand{\Author}{\begin{center}
   \large \bf
Guang-Liang Li${}^{a,b}$,  Xiaotian
Xu${}^{b,c,d}$, Kun Hao${}^{b,c,d}\footnote{Corresponding author:
haoke72@163.com}$, Pei Sun${}^{b,d,e}$, Junpeng Cao${}^{b,f,g,h}\footnote{Corresponding author:
junpengcao@iphy.ac.cn}$, Wen-Li
Yang${}^{b,c,d,e}\footnote{Corresponding author:
wlyang@nwu.edu.cn}$, Kangjie Shi${}^{c}$ and Yupeng
Wang${}^{b,f,i}$
 \end{center}}

\newcommand{\Address}{\begin{center}
${}^a$ Ministry of Education Key Laboratory for Nonequilibrium Synthesis and Modulation of Condensed Matter, School of
Physics, Xi'an Jiaotong University, Xi'an 710049, China\\
${}^b$ Peng Huanwu Center for Fundamental Theory, Xi'an 710127, China\\
${}^c$ Institute of Modern Physics, Northwest University, Xi'an 710127, China\\
${}^d$ Shaanxi Key Laboratory for Theoretical Physics Frontiers, Xi'an 710127, China\\
${}^e$ School of Physics, Northwest University, Xi'an 710127, China\\
${}^f$ Beijing National Laboratory for Condensed Matter Physics, Institute of Physics, Chinese Academy of Sciences, Beijing 100190, China\\
${}^g$ School of Physical Sciences, University of Chinese Academy of Sciences, Beijing 100049, China\\
${}^h$ Songshan Lake Materials Laboratory, Dongguan, Guangdong 523808, China \\
${}^i$ The Yangtze River Delta Physics Research Center, Liyang, Jiangsu, China\\
${}$ 
\end{center}}

\Title{Exact solution of the quantum integrable model associated with the twisted $D^{(2)}_3$ algebra}
\vspace{-0.32truecm}
\Author

\vspace{-0.32truecm}
\Address
%\vspace{0.2truecm}
\begin{abstract}

We generalize the nested off-diagonal Bethe ansatz method to study  the quantum chain associated with the twisted $D^{(2)}_3$ algebra (or the $D^{(2)}_3$ model) with either periodic or integrable open boundary conditions.
We obtain the intrinsic operator product identities among the fused transfer matrices and find a way to close the
recursive fusion relations, which makes it possible to determinate  eigenvalues of transfer matrices
with an arbitrary anisotropic parameter $\eta$. Based on them, and the asymptotic behaviors and values at certain points, we construct  eigenvalues of transfer matrices in terms of homogeneous
$T-Q$ relations for the periodic case and inhomogeneous ones for the open case with some off-diagonal boundary reflections.
The associated Bethe ansatz equations are also given. The method and results in this paper can be generalized to the $D^{(2)}_{n+1}$ model and other
high rank integrable models.

\vspace{0.5truecm} \noindent {\it PACS:} 75.10.Pq, 02.30.Ik, 71.10.Pm

\end{abstract}
\newpage
%%%%%%%%%%%%%%%%%%%%%%%%%%%%%%%%%%%%%%%%%%%%%%%%%%%%%%%%%%%%%%%
%                                                             %
%  1. Introduction                                            %
%                                                             %
%%%%%%%%%%%%%%%%%%%%%%%%%%%%%%%%%%%%%%%%%%%%%%%%%%%%%%%%%%%%%%%
\section{Introduction}
\label{intro} \setcounter{equation}{0}

One of the major achievements of one-dimensional quantum integrable systems is that it can supply us some believable results of
strong interacting quantum many-body systems. The coordinate \cite{c1,c2} and algebraic Bethe ansatz \cite{a1,a2,a3,a4,a5} as well as the $T-Q$ relations \cite{t1,t2,t3} are the powerful methods to obtain the exact solutions of the systems and
have made great achievements in the past several decades.
Based on the exact solution, many interesting physical concepts and mathematical structures in the quantum field theory, quantum group and quantum algebra are obtained \cite{p1}.

Recently, much attention has been paid on the investigation of quantum integrable systems with $U(1)$ symmetry broken,
because this kind of systems has many important applications in the open string, quantum magnetism and non-equilibrium statistical mechanics.
The typical models are the systems with twisted boundary conditions \cite{Yung95} or non-diagonal boundary magnetic fields \cite{3,4,5,6}.
Due to the fact that the traditional Bethe ansatz does not work, many interesting methods such as q-Qnsager alegebra \cite{Bas13}, separation of variables \cite{Skl95,Fra08, Fra11, Nic12}, off-diagonal Bethe ansatz (ODBA) \cite{cao13, cao13-1, wang15}, and modified algebraic Bethe ansatz \cite{Bel13, Bel15, Pim15, Ava15} were developed .

Motivated by the applications in AdS/CFT, string theory and conformal field theory, the
study of quantum integrable models with high rank Lie algebras becomes a very important issue \cite{NYRes,Bn,lima-1}. Many efforts have been made to investigate this kind of systems. For example,
the exact solutions of open boundary quantum integrable models associated with $A_n$ \cite{Cao14}, $B_2$ \cite{Li_119}, $C_n$ \cite{Li_221}, $D_3^{(1)}$ \cite{Li_319} and $A_2^{(2)}$ \cite{Hao14} Lie algebras are obtained.
Based on the $su(2)\times su(2)$ symmetry, some new integrable strongly correlated electronic systems are very recently  constructed \cite{s2}.

The $D_{n+1}^{(2)}$ is a typical twisted Lie affine algebra and the related quantum integrable models have been attracted many attentions  \cite{NYReshetikhin2,guan,5-1,lima-2,5-2,5-3,5-4}.
Reshetikhin obtained the Bethe ansatz solutions of the $D^{(2)}_{n+1}$ model with periodic boundary condition \cite{NYReshetikhin2}.
Martins and Guan studied the integrability of the $D^{(2)}_{n+1}$ model with  open boundary condition \cite{guan}.
Further, Nepomechie, Pimenta and Retore studied the integrable quantum group invariant and $D^{(2)}_{n+1}$ open spin chains, where an interesting result is that the $R$-matrix can be
constructed by two six-vertex $R$-matrices \cite{5-1}.
The integrable $D^{(2)}_{n+1}$ reflection matrices with quantum group symmetry and with other integrable boundary conditions were given in \cite{lima-2,5-2,5-3,Rob20,Rob21}. Another important progress is
the Bethe ansatz solution of the $D^{(2)}_{2}$ model \cite{Rob20,5-3,Rob21,5-4}, which has  application in the black hole theory.

In this paper, we study the quantum integrable spin chain associated with the twisted $D_3^{(2)}$. We generalize the nested ODBA method to the chain
with either the periodic or the open boundary condition.
By using the fusion technique \cite{Kul81,Kul86,Kar79,Kir86,Kir87,Mez92}, we systematically analyze the fusion structure of the system.
We provide a way to close the recursive fusion relations, which make the fusion relations can be used to construct the energy spectrum without any additional constraints.
We obtain the closed intrinsic operator product identities among the fused transfer matrices.
Based on them, and the asymptotic behaviors and values at certain points, we obtain
the eigenvalues of fused transfer matrices, which are expressed as the homogeneous $T-Q$ relations for the periodic case and inhomogeneous ones for the open case with
 off-diagonal reflection matrices. The associated Bethe ansatz equations are also given.

The plan of the paper is as follows.
In section 2, we study the $D^{(2)}_3$ spin chain with the periodic boundary condition.
The closed operator product identities among the fused transfer matrices are given. By constructing the homogeneous $T-Q$ relations,
we obtain the eigenvalues and Bethe ansatz equations of the system.
Section 3 is devoted to diagonalize the model with some non-diagonal boundary reflections. We obtain the recursive fusion relations,
the eigenvalues of transfer matrices in terms of inhomogeneous $T-Q$ relations,
and the Bethe ansatz equations. The summary of main results and some concluding remarks are presented in section 4. Some details deriving the fusions of the $R$-matrices and related $K$-matrices
are given in Appendices A and B.

%%%%%%%%%%%%%%%%%%%%%%%%%%%%%%%%%%%%%%%%%%%%%%%%%%%%%%%%%%%%%%%
%                                                             %
%  2. Transfer matrix                                         %
%                                                             %
%                                                             %
%                                                             %
%%%%%%%%%%%%%%%%%%%%%%%%%%%%%%%%%%%%%%%%%%%%%%%%%%%%%%%%%%%%%%%

\section{$D^{(2)}_3$ model with the periodic boundary condition}

\setcounter{equation}{0}

%\subsection{Vectorial $R$-matrix}

Throughout this paper, we adopt the standard notations. ${\rm\bf V}$ denotes a $n$-dimensional linear space
with the orthogonal basis $\{|i\rangle, i=1,2,\cdots,n\}$.
For any matrix $A\in {\rm End}({\rm\bf V})$, $A_j$ is an
embedding operator in the tensor space ${\rm\bf V}\otimes
{\rm\bf V}\otimes\cdots$, which acts as $A$ on the $j$-th space and as
identity on the other factor spaces. For any matrix $B\in {\rm End}({\rm\bf V}\otimes {\rm\bf V})$, $B_{ij}$ is an embedding
operator of $B$ in the tensor space, which acts as identity
on the factor spaces except for the $i$-th and $j$-th ones.

Let us consider the fundamental $R$-matrix associated with the vectorial representation of the twisted algebra $D_3^{(2)}$ given by
\bea R^{vv}_{12}(u)=U_1U_2 R_{12}(u) U_2^{-1}U_1^{-1}, \label{Rv}\eea
where $U$ is the gauge transformation  with the form of
 \bea
U=\left(\begin{array}{cccccc}\cosh\eta&0&0&0&0&0\\[6pt]
    0&\cosh\eta&0&0&0&0\\[6pt]
    0&0&\sqrt{\cosh 3\eta}&0&0&0\\[6pt]
    0&0&2\sinh^2\eta\sqrt{\cosh\eta}&-\cosh 2\eta\sqrt{\cosh\eta}&0&0\\[6pt]
    0&0&0&0&\cosh 2\eta&0\\[6pt]
0&0&0&0&0&\cosh 2\eta\end{array}\right),\label{Rd-s}\eea
and the matrix $R_{12}(u)$ is given by \cite{5-1}
\begin{eqnarray}
&&R_{12}(u)=e^{-2(u+3\eta)}\Big\{
\left(e^{2u}-e^{4\eta}\right)\left(e^{2u}-e^{8\eta}\right)\sum_{\alpha\neq
3,4}[e_1]^{\alpha}_{\alpha}\otimes [e_2]^{\alpha}_{\alpha}
+e^{2\eta}\left(e^{2u}-1\right)\left(e^{2u}-e^{8\eta}\right)
\nonumber\\[6pt]
&&\hspace{0.8truecm} \times \sum_{\substack{\alpha\neq\beta,\beta'\\\alpha\,\textrm{or}\,\beta\neq
3,4}}  [e_1]^{\alpha}_{\alpha}\otimes [e_2]^{\beta}_{\beta}-
\left(e^{4\eta}-1\right) \left(e^{2u}-e^{8\eta}\right)
\Big(\sum_{\substack{\alpha<\beta,\alpha\neq\beta' \\
\alpha,\beta\neq 3,4}}+e^{2u}\sum_{\substack{\alpha>\beta,\alpha\neq \beta'\\
\alpha,\beta\neq 3,4}}\Big)
 [e_1]^{\alpha}_{\beta}\otimes [e_2]^{\beta}_{\alpha}
\nonumber\\[6pt]
&& \hspace{0.8truecm}-\frac{1}{2} \left(e^{4\eta}-1\right)\left(e^{2u}-e^{8\eta}
\right) \Big[ \left(e^{u}+1\right)
 \Big(
\sum_{\alpha<3,\beta=3,4}+e^{u}\sum_{\alpha>4,\beta=3,4} \Big)
\no\\[6pt]
&&\hspace{0.8truecm} \times \Big( [e_1]^{\alpha}_{\beta}\otimes
[e_2]^{\beta}_{\alpha}+[e_1]^{\beta'}_{\alpha'}\otimes [e_2]^{\alpha'}_{\beta'} \Big)
+ \left(e^{u}-1\right)
 \Big(
-\sum_{\alpha<3,\beta=3,4}+e^{u}\sum_{\alpha>4,\beta=3,4} \Big)
\no\\
&&\hspace{0.8truecm} \times \left( [e_1]^{\alpha}_{\beta}\otimes
[e_2]^{\beta'}_{\alpha}+[e_1]^{\beta'}_{\alpha'}\otimes [e_2]^{\alpha'}_{\beta}
\right)\Big]+ \sum_{\alpha,\beta\neq
3,4}a^{\alpha}_{\beta}(u)[e_1]^{\alpha}_{\beta}\otimes [e_2]^{\alpha'}_{\beta'}+
 \frac{1}{2}\sum_{\alpha\neq 3,4,\beta=3,4}
\no\\[6pt]
&&\hspace{0.8truecm} \times \left[ b_\alpha^{+}(u) \left( [e_1]^{\alpha}_{\beta}\otimes
[e_2]^{\alpha'}_{\beta'}+[e_1]^{\beta'}_{\alpha'}\otimes [e_2]^{\beta}_{\alpha}
 \right)
+ b_\alpha^{-}(u) \left( [e_1]^{\alpha}_{\beta}\otimes
[e_2]^{\alpha'}_{\beta}+[e_1]^{\beta}_{\alpha'}\otimes [e_2]^{\beta}_{\alpha}
 \right)
\right]
\no\\[6pt]
&&\hspace{0.8truecm} +\sum_{\alpha=3,4} \left[c^{+}(u)[e_1]^{\alpha}_{\alpha}\otimes
[e_2]^{\alpha'}_{\alpha'}+ c^{-}(u)[e_1]^{\alpha}_{\alpha}\otimes
[e_2]^{\alpha}_{\alpha}\right.
\no\\[6pt]
&&\hspace{0.8truecm} \left. +\, d^{+}(u)[e_1]^{\alpha}_{\alpha'}\otimes [e_2]^{\alpha'}_{\alpha}+
d^{-}(u)[e_1]^{\alpha}_{\alpha'}\otimes [e_2]^{\alpha}_{\alpha'} \right]\Big\}.
\label{RmatD2}
\end{eqnarray}
Here $u$ is the spectral parameter and $\eta$ is the crossing or anisotropic parameter. The subscripts $\alpha,\beta=1,\cdots,6$, $\alpha'=7-\alpha$ and $\beta'=7-\beta$.
Each of matrices $[e_k]^{\alpha}_{\beta} (k=1,2) $ is the Weyl basis of the $6\times 6$ representation matrix of the $k$-th space. The coefficients   $a^{\alpha}_{\beta}(u)$  are
\begin{equation}
a^{\alpha}_{\beta}(u)=\begin{cases}
(e^{4\eta}e^{2u}-e^{8\eta})(e^{2u}-1),& \alpha=\beta,\\[6pt]
(e^{4\eta}-1)(e^{8\eta}e^{2\eta(\bar\alpha-\bar\beta)}(e^{2u}-1)-\delta_{\alpha\beta'}(e^{2u}-e^{8\eta})),& \alpha<\beta,\\[6pt]
(e^{4\eta}-1)e^{2u}(e^{2\eta(\bar\alpha-\bar\beta)}(e^{2u}-1)-\delta_{\alpha\beta'}(e^{2u}-e^{8\eta})),&
\alpha>\beta,
\end{cases}
\end{equation}
where $\alpha,\beta\neq 3,4$ is supposed and we have used the notation
\begin{equation}
\bar{\alpha}=\begin{cases}
\alpha+1, & 1\leq \alpha <3,\\[6pt]
\frac{7}{2}, & \alpha=3, 4,\\[6pt]
\alpha-1, & 4<\alpha \leq 6.
\end{cases}
\label{alphabarD2}
\end{equation}
The functions $b_{\alpha}^{\pm}(u)$, $c^{\pm}(u)$ and $d^{\pm}(u)$ are
\begin{eqnarray}
&&b_{\alpha}^{\pm}(u)=\begin{cases}
\pm e^{2\eta(\alpha-1/2)}(e^{4\eta}-1)(e^{2u}-1)(e^u\pm e^{4\eta}),& \alpha<3,\\[6pt]
e^{2\eta(\alpha-9/2)}(e^{4\eta}-1)(e^{2u}-1)e^u(e^u\pm e^{4\eta}),&
\alpha>4,
\end{cases} \no \\[6pt]
&&c^{\pm}(u)=\pm\frac{1}{2}(e^{4\eta}-1)(e^{4\eta}+1)e^u(e^u\mp
1)(e^u\pm e^{4\eta})+e^{2\eta}(e^{2u}-1)(e^{2u}-e^{8\eta}), \no \\[6pt]
&&d^{\pm}(u)=\pm\frac{1}{2}(e^{4\eta}-1)(e^{4\eta}-1)e^u(e^u\pm
1)(e^u\pm e^{4\eta}).
\end{eqnarray}
The $R$-matrix (\ref{Rv}) satisfies the properties
\begin{eqnarray}
\hspace{-0.8truecm}{\rm regularity}&:&R^{  vv}_{12}(0)=\rho_1(0)^{\frac{1}{2}}{\cal P}_{12},\\[4pt]
\hspace{-0.8truecm}{\rm unitarity}&:&R^{   vv}_{12}(u)R^{  vv}_{21}(-u)=\rho_1(u)=a_1(u)a_1(-u),\label{1-Unitarity}\\[4pt]
\hspace{-0.8truecm}{\rm crossing \,\,unitarity}&:&R^{
vv}_{12}(u)^{t_1}M_1R^{
vv}_{21}(-u+8\eta)^{t_1}M_1^{-1}=\rho_1(u-4\eta),\label{Crossing-symmetry}
\end{eqnarray}
where ${\cal P}_{12}$ is the permutation operator with the matrix
elements $[{\cal
P}_{12}]^{\alpha\gamma}_{\beta\delta}=\delta_{\alpha\delta}\delta_{\beta\gamma}$,
$t_k$ denotes the transposition in the $k$-th space,
$a_1(u)=4\sinh(u-4\eta)\,\sinh(u-2\eta)$, $R _{21}(u)={\cal P}_{12}R
_{12}(u){\cal P}_{12}$. The matrix $M$ reads \bea
M=diag(e^{6\eta},e^{2\eta},1,1,e^{-2\eta},e^{-6\eta}).\eea The
$R$-matrix (\ref{Rv}) also has the periodicity \bea &&R^{
vv}_{12}(u+i\pi)=\bar{V}_2R^{ vv}_{12}(u)\bar{V}_2^{-1},
\label{Rvpi}\eea where
\begin{eqnarray}
\bar{V}=\lt(\begin{array}{cccccc}\cosh 2\eta&&&&&\\
&\cosh 2\eta&&&&\\&&2\sinh^2\eta&-\sqrt{\frac{\cosh 3\eta}{\cosh\eta}}&&\\&&-\sqrt{\frac{\cosh 3\eta}{\cosh\eta}}&-2\sinh^2\eta&&\\&&&&\cosh 2\eta&\\
&&&&&\cosh 2\eta
\end{array}\rt).\label{vbar}
\end{eqnarray}
Besides the above properties, the $R$-matrix (\ref{Rv}) satisfies the Yang-Baxter equation
\begin{eqnarray}
R^{   vv}_{12}(u-v)R^{  vv}_{13}(u)R^{    vv}_{23}(v)=R^{
  vv}_{23}(v)R^{   vv}_{13}(u)R^{  vv}_{12}(u-v). \label{20190802-1}
\end{eqnarray}

The monodromy matrix of the system is constructed by the vectorial
$R$-matrix (\ref{Rv}) as
\bea
T_0^{  v}(u)=R^{    vv}_{01}(u-\theta_1)R^{
vv}_{02}(u-\theta_2)\cdots R^{
   vv}_{0N}(u-\theta_N), \label{Mon-1}
\eea
where the index $0$ indicates the auxiliary space and the indices $\{1, \cdots, N\}$ denote
the physical or quantum spaces, $N$ is the number of sites and $\{\theta_j|j=1, \cdots, N\}$ are the inhomogeneous parameters.
The monodromy matrix satisfies the Yang-Baxter relation
\bea
R^{  vv}_{12}(u-v) T_1^{  v}(u) T_2^{  v}(v) = T_2^{  v}(v) T_1^{  v}(u) R^{  vv}_{12}(u-v).\label{ybta2o}
\eea
The transfer matrix is defined as the partial trace of monodromy matrix in the auxiliary space
\bea
t^{(p)}(u)=tr_0 T_0^{  v}(u), \label{1117-1}
\eea
where the up-index $(p)$ of the transfer matrix $t^{(p)}(u)$ just stands for the periodic boundary condition (or the closed chain).
From the Yang-Baxter relation (\ref{ybta2o}), one can prove that the transfer matrices with different spectral parameters
commute with each other, i.e., $[t^{(p)}(u), t^{(p)}(v)]=0$. Therefore, $t^{(p)}(u)$ serves
as the generating functional of the conserved quantities of the
system. The Hamiltonian of the $D^{(2)}_3$ spin chain with the periodic boundary condition can be
given  in terms of the  transfer matrix (\ref{1117-1}) as
\begin{eqnarray}
H_p= \frac{\partial \ln t^{(p)}(u)}{\partial
u}|_{u=0,\{\theta_j\}=0}=\sum_{k=1}^{N}H_{k\,k+1},\label{haha-1}
\end{eqnarray}
with
\bea
H_{k\,k+1}= {\cal P}_{k\,k+1}\,\frac{\partial}{\partial u}R_{k\,k+1}(u)\left.\right|_{u=0}. \label{local-Ham}
\eea
The periodic boundary condition reads
\bea
H_{N\,N+1}=H_{N\,1}.\label{Periodic-boundary-condition}
\eea

\subsection{Operator product identities among the fused transfer matrices}

In order to obtain eigenvalues of the fundamental transfer matrix
$t^{(p)}(u)$, we need  further to introduce some  fused transfer
matrices \cite{wang15}. For the $D^{(2)}_3$ spin chain with the
periodic boundary condition, we introduce the fused monodromy
matrices via the fused $R$-matrices $R^{s_{\pm}v}_{\tilde{0}'j}(u)$ given by (\ref{rsp}) and (\ref{rsn}) in the
Appendix A,  \bea T_{\tilde{0}'}^{
\pm}(u)=R^{s_{\pm}v}_{\tilde{0}'1}(u-\theta_1)R^{
s_{\pm}v}_{\tilde{0}'2}(u-\theta_2)\cdots R^{
s_{\pm}v}_{\tilde{0}'N}(u-\theta_N).\label{Transfer-S-} \eea Here
and after, $\tilde{0}'=0'$ denotes the auxiliary space for the
spinorial representation $s_{+}$ and $\tilde{0}'={\bar 0}'$
denotes the auxiliary space for the spinorial representation
$s_{-}$. We note that the quantum spaces of $T_{{0}'}^{+}(u)$
and $T_{\bar{0}'}^{-}(u)$ are the same, which are also the quantum
spaces of $T_{0}(u)$. The fused $R$-matrices
$R^{s_{\pm}v}_{\tilde{0}'j}$ satisfy the Yang-Baxter relations
\bea
&&R^{  vs_{\pm}}_{0\tilde{0}'}(u-v) T_0(u) T_{\tilde{0}'}^{\pm}(v) = T_{\tilde{0}'}^{\pm}(v)T_0(u)  R^{  vs_{\pm}}_{0\tilde{0}'}(u-v),  \label{yang-1}\\
&&R^{s_+s_-}_{0'\bar{0}'}(u-v) T_{0'}^{+}(u) T_{\bar{0}'}^{-}(v) =
T_{\bar{0}'}^{-}(v)T_{0'}^{+}(u)  R^{s_+s_-}_{0'\bar{0}'}(u-v),
\label{y1bta2o} \eea where $R^{s_+s_-}_{0'\bar{0}'}(u)$ is defined
by Eq.(\ref{Rsspm}). Taking the partial trace in the auxiliary
spaces, we obtain the fused transfer matrices \bea
 t^{(p)}_{\pm}(u)=tr_{\tilde{0}'} T_{\tilde{0}'}^{\pm}(u).
\eea
From the Yang-Baxter relation (\ref{y1bta2o}), we can prove that the transfer matrices $ t^{(p)}(u)$ and $ t^{(p)}_{\pm}(u)$ commute with each other, i.e.,
\bea
[t^{(p)}(u), t^{(p)}_{\pm}(u)]=[t^{(p)}_+(u), t^{(p)}_{-}(u)]=0.
\eea
Thus they have the common eigenstates.

From the Yang-Baxter relations (\ref{ybta2o}), (\ref{yang-1}) and (\ref{y1bta2o})
at certain points and using the properties of projectors, we
obtain \bea &&T_1(\theta_j)\,T_2(\theta_j+4\eta)=
P^{{  vv}(1) }_{21}\,T_1(\theta_j)\,T_2(\theta_j+4\eta),  \no \\
&&T_2(\theta_j)\,T_1(\theta_j+2\eta+i\pi)=P^{{  vv} (16) }_{12}\,T_2(\theta_j)\,T_1(\theta_j+2\eta+i\pi),\no \\
&&T_2(\theta_j)\, T_{\tilde{1}'}^{\pm}(\theta_j+3\eta+i\pi)=
P_{\tilde{1}'2}^{(\pm)}\,T_2(\theta_j)T_{\tilde{1}'}^{\pm}(\theta_j+3\eta+i\pi),\quad
j=1,\cdots,N, \label{opr-1}
\eea
where the projectors: $P^{{  vv}(1) }_{21}$, $P^{{  vv} (16) }_{12}$, $P_{\tilde{1}'2}^{(\pm)}$ are given by (\ref{1a1}),
(\ref{16-dim}), (\ref{4-dim-1}) and (\ref{4-dim-2}). By using the fusion identities
(\ref{sv-2-1}), (\ref{ai}), (\ref{hhhgg-1}) and (\ref{uf-12}), we
obtain the fusion identities:
\bea &&P^{{  vv}(1)
}_{21}T_1(u)\,T_2(u+4\eta)P^{{ vv}(1) }_{21}= P^{{  vv}(1)
}_{21}\prod_{j=1}^N
a_1(u-\theta_j)e_1(u-\theta_j+4\eta)\times {\rm id}, \no \\
&&P^{{  vv} (16) }_{12}T_2(u)\,T_1(u+2\eta+i\pi)P^{{  vv} (16)
}_{12}\no \\
&&\qquad\qquad=\prod_{j=1}^N
\tilde{\rho}_0(u-\theta_j)S_{1'2'}\,T^+_{1'}(u+\eta+i\pi)\,T^-_{2'}(u+\eta+i\pi)S_{1'2'}^{-1},\no \\
&&P_{1'2}^{(+)}T_2(u)\, T_{1'}^{ +}(u+3\eta+i\pi)P_{1'2}^{(+)}=
\prod_{j=1}^N
\tilde{\rho}_0(u-\theta_j)\, T_{\langle 1'2\rangle }^{ -}(u+\eta+i\pi),\no \\
&&P_{\bar{1}'2}^{(-)}T_2(u)\, T_{\bar{1}'}^{
-}(u+3\eta+i\pi)P_{\bar{1}'2}^{(-)}= \prod_{j=1}^N
\tilde{\rho}_0(u-\theta_j)\, \tilde{S}_{\langle
\bar{1}'2\rangle}T_{\langle \bar{1}'2\rangle }^{
+}(u+\eta+i\pi)\tilde{S}_{\langle
\bar{1}'2\rangle}^{-1}.\label{opr-11}
 \eea
Taking the partial trace of Eq.(\ref{opr-11}) in the auxiliary spaces and using
the relation (\ref{opr-1}), we obtain the operator product identities \bea &&
t^{(p)}(\theta_j)\,t^{(p)}(\theta_j+4\eta)=\prod_{l=1}^N
a_1(\theta_j-\theta_l)e_1(\theta_j-\theta_l+4\eta)\times {\rm id}, \no  \\
&& t^{(p)}(\theta_j)\,t^{(p)}(\theta_j+2\eta+i\pi)= \prod_{l=1}^N
\tilde{\rho}_0(\theta_j-\theta_l)\,t^{(p)}_+(\theta_j+\eta+i\pi)\,t^{(p)}_{-}(\theta_j+\eta+i\pi),\no  \\
&&t^{(p)}(\theta_j)\,t^{(p)}_{\pm}(\theta_j+3\eta+i\pi)=\prod_{l=1}^N
\tilde{\rho}_0(\theta_j-\theta_l)\,t^{(p)}_{\mp}(\theta_j+\eta+i\pi),\quad j=1,\cdots,N.\label{fut1p-5}\eea

Next, we consider the asymptotic behaviors of fused transfer matrices.
According to the definitions, the direct calculation gives
\bea && t^{ (p)}(u)|_{u\rightarrow \pm\infty}=e^{\pm (2Nu-\sum_{j=1}^N\theta_j)}
\sum_{\alpha=1}^6[T_{\pm}^{v}]^{\alpha}_{\alpha}+\cdots,\no \\
&& t^{(p)}_{\pm}(u)|_{u\rightarrow \pm\infty}=e^{\pm(
Nu-\sum_{j=1}^N\theta_j)}
\sum_{\alpha=1}^4[T_{\pm}^{s_\pm}]^{\alpha}_{\alpha} +\cdots,
\label{fuwwwtpl-7} \eea where $\sum_{\alpha=1}^6[T_{\pm}^{v}]^{\alpha}_{\alpha}$
and $\sum_{\alpha=1}^4[T_{\pm}^{s_\pm}]^{\alpha}_{\alpha}$ are the conserved quantities acting
on the quantum space ${\rm\bf V}\otimes {\rm\bf
V}\otimes\cdots\otimes{\rm\bf V}$. The related operators are defined as  \bea
&&[T_{\pm}^{v}]^{\alpha}_{\beta}=\sum_{\{\delta_i\}=1,\{\gamma_i\}=1}^6
[R^{vv(\pm)}_{01}]^{\alpha \gamma_1}_{\alpha_1 \delta_1}
[R^{vv(\pm)}_{02}]^{\alpha_1 \gamma_2}_{\alpha_2 \delta_2} \cdots
[R^{vv(\pm)}_{0N}]^{\alpha_{N-1} \gamma_N}_{\beta\quad\;\;\, \delta_N},\no\\
&&[T_{\pm}^{s_\pm}]^{\alpha}_{\beta}=\sum_{\{\delta_i\}=1,\{\gamma_i\}=1}^6[R^{s_\pm
v(\pm)}_{\tilde{0}'1}]^{\alpha \gamma_1}_{\alpha_1 \delta_1}
[R^{s_\pm v(\pm)}_{\tilde{0}'2}]^{\alpha_1 \gamma_2}_{\alpha_2
\delta_2} \cdots [R^{s_\pm v(\pm)}_{\tilde{0}'N}]^{\alpha_{N-1}
\gamma_N}_{\beta\quad\;\;\, \delta_N}. \label{tjj-1}\eea Here the
repeated indicators should be summarized. $R^{vv(\pm)}_{0j}$ and
$R^{s_{\pm}v(\pm)}_{\tilde{0}'j}$ are the leading terms of $e^{\mp
2u}R^{vv}_{0j}(u)$ and $e^{\mp u}R^{s_{\pm}v}_{\tilde{0}'j}(u)$
 with $u\rightarrow\pm\infty$, respectively. From the
direct calculation, we find that the eigenvalues of conserved
quantities $\sum_{\alpha=1}^6[T_{\pm}^{v}]^{\alpha}_{\alpha}$ and
$\sum_{\alpha=1}^4[T_{\pm}^{s_\pm}]^{\alpha}_{\alpha}$ can be
expressed by two quantum numbers $m_1$ and $m_2$ as
$2[1+\cosh(2m_1\eta)+\cosh(2m_2\eta)]e^{\mp 4N\eta}$ and
 $2\{\cosh[(m_1+m_2)\eta]
+\cosh[(m_1-m_2)\eta]\}e^{\mp 2N\eta}$, respectively, where
$m_1\in [0,N]$ and $0 \leq m_2\leq N-m_1$. Then the asymptotic behaviors of
fused transfer matrices read \bea &&\hspace{-1.4cm} t^{ (p)}(u)|_{u\rightarrow
\pm\infty}= 2[1+\cosh(2m_1\eta)
+\cosh(2m_2\eta)]e^{\pm (2Nu-\sum_{j=1}^N\theta_j-4N\eta)} +\cdots,\no \\
&&\hspace{-1.4cm} t^{(p)}_{\pm}(u)|_{u\rightarrow \pm\infty}=
2\{\cosh[(m_1+m_2)\eta]
+\cosh[(m_1-m_2)\eta]\} e^{\pm
(Nu-\sum_{j=1}^N\theta_j-2N\eta)} +\cdots.\label{fuwwsdsdw2tpl-7} \eea

Acting the transfer matrices on the common eigenstate, we obtain the corresponding eigenvalues. Denote the eigenvalues of
$t^{(p)}(u)$ and $t^{(p)}_{\pm}(u)$ as $\Lambda^{(p)}(u)$ and
$\Lambda^{(p)}_{\pm}(u)$, respectively. As mentioned previously, the eigenvalues $\Lambda^{(p)}(u)$ and
$\Lambda^{(p)}_{\pm}(u)$ are the trigonometric polynomials of $u$ with degrees
$2N$ and $N$, respectively. Therefore, we need $4N+3$ conditions to
determine the values of $\Lambda^{(p)}(u)$ and
$\Lambda^{(p)}_{\pm}(u)$.

From the operator product identities (\ref{fut1p-5}), we have the
functional relations among the eigenvalues \bea &&
\Lambda^{(p)}(\theta_j)\,\Lambda^{(p)}(\theta_j+4\eta)=\prod_{l=1}^N
a_1(\theta_j-\theta_l)e_1(\theta_j-\theta_l+4\eta),\no  \\
&& \Lambda^{(p)}(\theta_j)\,\Lambda^{(p)}(\theta_j+2\eta+i\pi)=
\prod_{l=1}^N
\tilde{\rho}_0(\theta_j-\theta_l)\,\Lambda^{(p)}_+(\theta_j+\eta+i\pi)
\,\Lambda^{(p)}_{-}(\theta_j+\eta+i\pi),\no  \\
&&\Lambda^{(p)}(\theta_j)\Lambda^{(p)}_{\pm}(\theta_j+3\eta+i\pi)=\prod_{l=1}^N
\tilde{\rho}_0(\theta_j-\theta_l)\,\Lambda^{(p)}_{\mp}(\theta_j+\eta+i\pi),\quad j=1,\cdots,N.\label{l1-4}\eea
The corresponding asymptotic behaviors are
\bea && \Lambda^{
(p)}(u)|_{u\rightarrow \pm\infty}= 2[1+\cosh(2m_1\eta)
+\cosh(2m_2\eta)]e^{\pm (2Nu-\sum_{j=1}^N\theta_j-4N\eta)} +\cdots,\no \\
&& \Lambda^{(p)}_{\pm}(u)|_{u\rightarrow \pm\infty}=
2\{\cosh[(m_1+m_2)\eta]
+\cosh[(m_1-m_2)\eta]\}\no \\
&&\hspace{3cm} \times e^{\pm
(Nu-\sum_{j=1}^N\theta_j-2N\eta)} +\cdots.\label{fuwww2tpl-7} \eea
Then we
arrive at that $4N$ functional relations (\ref{l1-4}) together
with $6$ asymptotic behaviors (\ref{fuwww2tpl-7}) give us
sufficient conditions to determine the eigenvalues of transfer matrices.

\subsection{$T-Q$ relations for eigenvalues}

The function relations (\ref{l1-4}) and asymptotic behaviors (\ref{fuwww2tpl-7}) allow us to parameterize the eigenvalues $ \Lambda^{(p)}(u)$ and $\Lambda^{(p)}_{\pm}(u)$ in terms of the $T-Q$ relations as \bea
\Lambda^{(p)}(u)&=&\prod_{j=1}^N
a_1(u-\theta_j)\frac{Q_{p}^{(1)}(u+2\eta)}{Q_{p}^{(1)}(u)}
+\prod_{j=1}^Nb_1(u-\theta_j)\Big\{\frac{Q_{p}^{(2)}(u+2\eta)Q_{p}^{(3)}(u-2\eta)}{Q_{p}^{(2)}(u)Q_{p}^{(3)}(u)}\no\\[4pt]
&&
+\frac{Q_{p}^{(1)}(u-2\eta)
Q_{p}^{(2)}(u+2\eta)Q_{p}^{(3)}(u+2\eta)}{Q_{p}^{(1)}(u)Q_{p}^{(2)}(u)Q_{p}^{(3)}(u)}+\frac{Q_{p}^{(2)}(u-2\eta)Q_{p}^{(3)}(u+2\eta)}{Q_{p}^{(2)}(u)Q_{p}^{(3)}(u)}
\no\\[4pt]
&&+\frac{Q_{p}^{(1)}(u)Q_{p}^{(2)}(u-2\eta)Q_{p}^{(3)}(u-2\eta)}{Q_{p}^{(1)}(u-2\eta)Q_{p}^{(2)}(u)Q_{p}^{(3)}(u)}\Big\}
+\prod_{j=1}^N e_1(u-\theta_j)\,
\frac{Q_{p}^{(1)}(u-4\eta)}{Q_{p}^{(1)}(u-2\eta)}, \label{T-Q-Periodic1}\\
\Lambda^{(p)}_+(u)&=&\prod_{j=1}^N
a_2(u-\theta_j)\left[\frac{Q^{(2)}_p(u+3\eta)}{Q^{(2)}_p(u+\eta)}
+\frac{Q^{(1)}_p(u+\eta)Q^{(2)}_p(u-\eta)}{Q^{(1)}_p(u-\eta)Q^{(2)}_p(u+\eta)}\right]\no\\
&&+\prod_{j=1}^N
b_2(u-\theta_j)\left[\frac{Q^{(3)}_p(u-3\eta)}{Q^{(3)}_p(u-\eta)}
+\frac{Q^{(1)}_p(u-3\eta)Q^{(3)}_p(u+\eta)}{Q^{(1)}_p(u-\eta)Q^{(3)}_p(u-\eta)}\right], \label{T-Q-Periodic2}\\
\Lambda^{(p)}_-(u)&=&\prod_{j=1}^N
a_2(u-\theta_j)\left[\frac{Q^{(3)}_p(u+3\eta)}{Q^{(3)}_p(u+\eta)}
+\frac{Q^{(1)}_p(u+\eta)Q^{(3)}_p(u-\eta)}{Q^{(1)}_p(u-\eta)Q^{(3)}_p(u+\eta)}\right]\no\\
&&+\prod_{j=1}^N
b_2(u-\theta_j)\left[\frac{Q^{(2)}_p(u-3\eta)}{Q^{(2)}_p(u-\eta)}
+\frac{Q^{(1)}_p(u-3\eta)Q^{(2)}_p(u+\eta)}{Q^{(1)}_p(u-\eta)Q^{(2)}_p(u-\eta)}\right],
\label{T-Q-Periodic} \eea
where
\bea
&&Q_{p}^{(1)}(u)=\prod_{k=1}^{L_1}\sinh(u-\mu_k^{(1)}-\eta),\;\;
Q_{p}^{(2)}(u)=Q_{p}^{(3)}(u-i\pi)=\prod_{l=1}^{L_2}\sinh\frac
12(u-\mu_l^{(2)}-2\eta),\no
\\&& b_1(u)=4\sinh u \sinh(u-4\eta), \;\; b_2(u)=2\sinh(u-\eta), \;\; a_2(u)=2\sinh(u-3\eta), \eea
$L_1$ is the number of Bethe roots $\{\mu^{(1)}_k\}$ and $L_2$ is the number of Bethe roots $\{\mu^{(2)}_l\}$.

Because the eigenvalues $\Lambda^{(p)}(u)$ and $\Lambda_{\pm}^{(p)}(u)$ are the polynomials, the regularity analyses of the
$T-Q$ relations (\ref{T-Q-Periodic1})-(\ref{T-Q-Periodic}) lead to that the
Bethe roots $\{\mu^{(1)}_k\}$ and $\{\mu^{(2)}_l\}$ should satisfy the Bethe ansatz
equations (BAEs) \bea &&
\frac{Q_{p}^{(1)}(\mu_k^{(1)}+3\eta)Q_{p}^{(2)}(\mu_k^{(1)}+\eta)Q_{p}^{(3)}(\mu_k^{(1)}+\eta)}
{Q_{p}^{(1)}(\mu_k^{(1)}-\eta)Q_{p}^{(2)}(\mu_k^{(1)}+3\eta)Q_{p}^{(3)}(\mu_k^{(1)}+3\eta)}
=-\prod_{j=1}^N \frac{\sinh(\mu_k^{(1)}+\eta-\theta_j)
}{\sinh(\mu_k^{(1)}-\eta-\theta_j)}, \no \\[6pt]&&
\qquad\qquad\qquad  k=1,\cdots, L_1, \label{BAEs-31} \\[6pt]
&&\frac{Q_{p}^{(1)}(\mu_l^{(2)})Q_{p}^{(2)}(\mu_l^{(2)}+4\eta)}
{Q_{p}^{(1)}(\mu_l^{(2)}+2\eta)Q_{p}^{(2)}(\mu_l^{(2)})} =-1,
\quad l=1,\cdots, L_2,\label{BAEs-3} \eea
where $L_1\leq N$ and $L_2\leq L_1$.

Some remarks are in order. We note that the BAEs (\ref{BAEs-31})
and (\ref{BAEs-3}) are homogeneous. This is because  the
periodic boundary condition does not break the $U(1)$ symmetry of
the system. The BAEs obtained from the regularity of
$\Lambda^{(p)}(u)$ are the same as those obtained from the
regularities of $\Lambda^{(p)}_{\pm}(u)$. From the asymptotic
behaviors of  $\Lambda^{(p)}(u)$ and $\Lambda_{\pm}^{(p)}(u)$, we
know that the quantum numbers $m_1$ and $m_2$ characterizing
the conserved quantities
$\sum_{\alpha=1}^6[T_{\pm}^{v}]^{\alpha}_{\alpha}$ and
$\sum_{\alpha=1}^4[T_{\pm}^{s_\pm}]^{\alpha}_{\alpha}$ are related
with the numbers of Bethe roots as
\begin{eqnarray}
m_1=N-L_1, \quad m_2=L_1-L_2. \label{m}
\end{eqnarray}
The existence of two good quantum numbers consists with the fact that there are two sets of homogeneous BAEs.
It is easy to check that $\Lambda^{(p)} (u)$ and
$\Lambda_{\pm}^{(p)}(u)$ satisfy the functional relations
(\ref{l1-4}) and the asymptotic behaviors (\ref{fuwww2tpl-7}).
Therefore, we conclude that $\Lambda^{(p)}(u)$ and
$\Lambda_{\pm}^{(p)}(u)$ are the eigenvalues of the transfer
matrices $t^{(p)}(u)$ and $t^{(p)}_{\pm}(u)$, respectively. We should note that the $T-Q$ relations
(\ref{T-Q-Periodic1})-(\ref{T-Q-Periodic}) and associated BAEs  (\ref{BAEs-31})-(\ref{BAEs-3}) have the well-defined homogeneous limit.
These results with the constraint $\{\theta_j\}=0$  coincide
with the previous results \cite{NYRes, NYReshetikhin2}.

The eigenvalues of the Hamiltonian (\ref{haha-1}) can be obtained by the $\Lambda^{(p)}(u)$ as
\begin{eqnarray}
E_p= \frac{\partial \ln \Lambda^{(p)}(u)}{\partial
u}|_{u=0,\{\theta_j\}=0}.
\end{eqnarray}

\section{$D^{(2)}_3$ model with non-diagonal boundary condition}
\setcounter{equation}{0}

In this section, we study the system with general integrable open boundary condition.
The boundary reflection at one side is quantified by the reflection matrix $K^{v}(u)$ which
satisfies the reflection equation
\begin{equation}
 R^{   vv}_{12}(u-v){K^{  v}_{  1}}(u)R^{   vv}_{21}(u+v) {K^{   v}_{2}}(v)=
 {K^{   v}_{2}}(v)R^{   vv}_{12}(u+v){K^{   v}_{1}}(u)R^{   vv}_{21}(u-v).
 \label{r1}
 \end{equation}
The boundary reflection at the other side is described by the dual reflection matrix $\bar K^{v}(u)$, which
satisfies the dual reflection equation
\begin{eqnarray}
 &&R^{   vv}_{12}(-u+v){\bar{K}^{   v}_{1}}(u)M_1^{-1}R^{  vv}_{21}
 (-u-v+8\eta)M_1{\bar{K}^{   v}_{2}}(v)\nonumber\\[4pt]
&&\qquad\qquad\quad\quad={\bar{K}^{   v}_{2}}(v)M_1R^{
vv}_{12}(-u-v+8\eta)M_1^{-1} {{\bar{K}}^{ v}_{1}}(u)R^{
vv}_{21}(-u+v).
 \label{r2}
 \end{eqnarray}
In the open boundary condition, besides the monodromy matrix $T^{v}_0(u)$ given by (\ref{Mon-1}), we should
also consider the reflecting monodromy matrix
\begin{eqnarray}
\hat{T}_0^{  v} (u)=R_{N0}^{  vv}(u+\theta_N)\cdots R_{20}^{  vv}(u+\theta_{2}) R_{10}^{  vv}(u+\theta_1),\label{Tt11}
\end{eqnarray}
which satisfies the Yang-Baxter relation
\begin{eqnarray}
R_{ 21}^{  vv} (u-v) \hat T_{1}^{  v}(u) \hat T_2^{  v}(v)=\hat  T_2^{  v}(v) \hat T_{ 1}^{  v}(u) R_{21}^{  vv} (u-v).\label{haishi0}
\end{eqnarray}
The transfer matrix $t(u)$ of the model with boundary reflections is defined as \cite{4}
\begin{equation}
t(u)= tr_0 \{\bar K_0^{  v }(u)T_0^{  v} (u) K^{  v }_0(u)\hat{T}^{  v}_0 (u)\}. \label{trweweu1110}
\end{equation}
From the Yang-Baxter relations (\ref{ybta2o}), (\ref{haishi0}),
reflection equation (\ref{r1}) and dual one (\ref{r2}), one can
prove that the transfer matrices with different spectral
parameters commute with each other, $[t(u), t(v)]=0$. Therefore,
$t(u)$ serves as the generating functional of all the conserved
quantities of the system. The Hamiltonian is constructed by taking
the derivative of the logarithm of the transfer
matrix\footnote{It should be remarked that one can define the Hamiltonian by (\ref{hh}) for the case of
$t(0)\ne 0$ (i.e., $tr_0 \bar K^{v}_0(0)\ne 0$), however for the case of $t(0)=0$ (i.e., $tr_0 \bar K^{v}_0(0)=0$) one
needs to adopt other way \cite{Rob21} instead to construct a meaningful Hamiltonian.  For the $K$-matrices (\ref{K-matrix-VV})-(\ref{ksk111}) that  we consider in the following parts of the paper, $tr_0 \bar K^{v}_0(0)\ne 0$ and (\ref{hh}) gives
a well-defined Hamiltonian.}
\begin{eqnarray}
H&=&\frac{1}{2}\frac{\partial \ln t(u)}{\partial u}|_{u=0,\{\theta_j\}=0}\nonumber \\
&=& \sum^{N-1}_{k=1}H_{k
k+1}+\frac{{K^{v}_N}(0)'}{2{K^{v}_N}(0)}+\frac{ tr_0 \{\bar
K^{v}_0(0)H_{10}\}}{tr_0 \bar K^{v}_0(0)}+{\rm constant},
\label{hh}
\end{eqnarray} where $H_{k k+1}$ is given by (\ref{local-Ham}).

\subsection{Reflection matrix}

In this paper, we consider the integrable open boundary condition
where the reflection matrices have the non-diagonal elements,
which break the $U(1)$-symmetry of the system. The non-diagonal
reflection matrix for $D_{n+1}^{(2)}$ vertex model has been
constructed by Malara et al \cite{lima-2} and Nepomechie et al
\cite{5-2}. Without losing generality, we here chose a reflection
matrix $K^{v}(u)$ with off-diagonal elements to demonstrate how
our method works, namely\footnote{The $K$-matrix $K^{v}(u)$ given by (\ref{K-matrix-VV})-(\ref{K-matrix-VV-1}) is one of special case of those obtained in
\cite{lima-2} after a gauge transformation.},
\bea
K^{v}(u)=\left(\begin{array}{cccccc}k_1(u) &0&0&0&k_4(u)&0\\[6pt]
    0&k_1(u)&0&0&0&-k_4(u)\\[6pt]
    0&0&k_2(u)&0&0&0\\[6pt]
    0&0&0&k_2(u)&0&0\\[6pt]
    k_5(u)&0&0&0&k_3(u) &0\\[6pt]
0&-k_5(u)&0&0&0&k_3(u)
\end{array}\right),
\label{K-matrix-VV}\eea  where the non-zero matrix elements are
 \bea&&k_1(u)=e^{-u}\cosh\eta,\quad
k_2(u)=\cosh(u+\eta),\quad k_3(u)=e^{u}\cosh\eta,\no\\[6pt]
 &&k_4(u)=-c\, {\sinh u}{\cosh 2\eta},\quad
k_5(u)=\frac{\sinh u}{c\cosh 2\eta}, \label{K-matrix-VV-1}\eea
and $c$ is a free boundary parameter.
The dual reflection matrix $\bar K^{v}(u)$ is obtained by the mapping
\begin{equation}
\bar K^{   v}(u)=M K^{ v}(-u+4\eta)|_{c\rightarrow\, c'},
\label{ksk111}
\end{equation}
where $c'$ is the boundary parameters at the other side. For a
generic choice of the  boundary parameters $\{c,\,c'\}$, it is
easy to check that $[K^{v}(u),\,\bar K^{v}(v)]\neq 0$, which
implies that the matrices $K^{v}(u)$ and $\bar K^{v}(u)$ cannot be
diagonalized simultaneously. Thus the $U(1)$ symmetry of the system is broken while the integrability is still held.

Substituting the expressions of $R$-matrix (\ref{Rv}) and reflection matrices (\ref{K-matrix-VV}) and (\ref{ksk111}) into (\ref{hh}),
we obtain the integrable Hamiltonian of $D^{(2)}_3$ model with the non-diagonal boundary reflections given by (\ref{K-matrix-VV}) and (\ref{ksk111}).

\subsection{Operators product relations}

Similarly as the periodic case in the previous section, we need further to introduce some fused transfer matrices (see below (\ref{transfer-open})) besides the fundamental one $t(u)$.  Due to the boundary reflection, besides the fused monodromy matrices ${T}_{ {\tilde{0}}'}^{\pm}(u)$ given by (\ref{Transfer-S-}), we should define the reflecting
fused monodromy matrices $\hat{T}_{ {\tilde{0}}'}^{\pm}(u)$ in
terms of fused $R$-matrix $R^{vs_{\pm}}_{j {\tilde{0}}'}(u)$
as
\begin{eqnarray}
\hat{T}_{ {\tilde{0}}'}^{ \pm}(u)=R^{    vs_{\pm}}_{N
{\tilde{0}'}}(u+\theta_N)\cdots R^{ vs_{\pm}}_{2
{\tilde{0}'}}(u+\theta_{2}) R^{
   vs_{\pm}}_{1 {\tilde{0}'}}(u+\theta_1),\label{M2on-2}
\end{eqnarray}
which satisfy the Yang-Baxter relations
\begin{eqnarray}
&&R_{0\tilde{0}'}^{ vs_{\pm}} (u-v)  \hat T_{\tilde{0}'}^{
\pm}(v)\hat T_{0}(u)
=\hat T_{0}(u)\hat  T_{\tilde{0}'}^{  \pm}(v)  R_{0\tilde{0}'}^{ vs_{\pm}} (u-v), \label{haish1i01} \\
&&R_{0'\bar{0}'}^{s_+ s_{-}} (u-v)  \hat T_{\bar{0}'}^{
s_-}(v)\hat T_{0'}^{s_+}(u)=\hat T_{0'}^{s_+}(u)\hat
T_{\bar{0}'}^{s_-}(v)  R_{0'\bar{0}'}^{ s_+s_{-}}
(u-v).\label{haish1i0}
\end{eqnarray}
The fused transfer matrices are constructed as \bea
t_{\pm}(u)=tr_{\tilde{0}'}\{\bar{K}^{s_{\pm}}_{\tilde{0}'}(u)
T_{\tilde{0}'}^{\pm}(u)K^{
s_{\pm}}_{\tilde{0}'}(u)\hat{T}_{\tilde{0}'}^{\pm}(u)\},\label{transfer-open}
\eea
where the fused K-matrices $K^{s_{\pm}}_{\tilde{0}'}(u)$ and $\bar{K}^{s_{\pm}}_{\tilde{0}'}(u)$ are given by (\ref{K-matrix-1}) and (\ref{K-matrix-spinor-1}).
From
the Yang-Baxter relations (\ref{y1bta2o}),
(\ref{haish1i01})-(\ref{haish1i0}) and reflection equations
(\ref{1r2})-(\ref{1dr3}), one can prove that the transfer matrices
$t(u)$ and $t_{\pm}(u)$ commute with each other,
\bea
[t(u),\, t_{\pm}(v)]=[t_+(u),\, t_{-}(v)]=[t_{\pm}(u),\,t_{\pm}(v)]=0.
\eea
Thus $t(u)$ and $t_{\pm}(u)$ have the common eigenstates.

Considering the Yang-Baxter relations (\ref{haishi0}) at the points of $u=-\theta_j$, $v=\{-\theta_j+4\eta, -\theta_j+2\eta+i\pi\}$,
(\ref{haish1i01}) at the points of $u=-\theta_j$, $v=-\theta_j+3\eta+i\pi$
and using the properties of projector, we obtain \bea
&&\hat{T}_1(-\theta_j)\,\hat{T}_2(-\theta_j+4\eta)=
P^{{  vv}(1) }_{12}\,\hat{T}_1(-\theta_j)\,\hat{T}_2(-\theta_j+4\eta), \no \\
&&\hat{T}_2(-\theta_j)\,\hat{T}_1(-\theta_j+2\eta+i\pi)=
P_{21}^{{  vv}(16) }\,\hat{T}_2(-\theta_j)\,\hat{T}_1(-\theta_j+2\eta+i\pi),\no \\
&&\hat{T}_2(-\theta_j)\,
\hat{T}_{\tilde{1}'}^{\pm}(-\theta_j+3\eta+i\pi)=
P_{2\tilde{1}'}^{(\pm)}\,\hat{T}_2(-\theta_j)\,
\hat{T}_{\tilde{1}'}^{ \pm}(-\theta_j+3\eta+i\pi), \label{op22r-1}
\eea
where $j=1,\cdots, N$, and the projectors: $P^{{  vv}(1) }_{21}$, $P^{{  vv} (16) }_{12}$, $P_{\tilde{1}'2}^{(\pm)}$ are given by (\ref{a1}),
(\ref{16-dim}), (\ref{4-dim-1}) and (\ref{4-dim-2}). Taking the fusion of reflecting monodromy matrices $\hat{T}(u)$, $\hat{T}^\pm(u)$ with the projectors $P^{{ vv}(1) }_{12}$, $P_{12}^{{  vv}(16) }$,
$P_{2\tilde{1}'}^{(\pm)}$ and using Eqs.(\ref{hhgg-1}),
(\ref{fu-12}), (\ref{sv-2}), (\ref{sv-1111}), we obtain the fusion
identities \bea &&P^{{ vv}(1)
}_{12}\hat{T}_1(u)\,\hat{T}_2(u+4\eta)P^{{ vv}(1) }_{12}= P^{{
vv}(1) }_{12}\prod_{j=1}^N
a(u+\theta_j)e(u+\theta_j+4\eta)\times {\rm id}, \no \\
&&P^{{  vv}(16) }_{21}\hat{T}_2(u)\,\hat{T}_1(u+2\eta+i\pi)P^{{
vv}(16) }_{21}\no \\
&& \qquad =\prod_{j=1}^N
\tilde{\rho}_0(u+\theta_j)\bar{S}_{1'\bar{2}'}\,\hat{T}^+_{1'}(u+\eta+i\pi)\,\hat{T}^-_{\bar{2}'}
(u+\eta+i\pi)\bar{S}_{1'\bar{2}'}^{-1},\no \\
&&P_{21'}^{(+)}\hat{T}_2(u)\, \hat{T}_{1'}^{
+}(u+3\eta+i\pi)P_{21'}^{(+)}= \prod_{j=1}^N
\tilde{\rho}_0(u+\theta_j)\, \hat{T}_{\langle 1'2\rangle }^{ -}(u+\eta+i\pi),\no \\
&&P_{2\bar{1}'}^{(-)}\hat{T}_2(u)\, \hat{T}_{\bar{1}'}^{
-}(u+3\eta+i\pi)P_{21'}^{(-)}= \prod_{j=1}^N
\tilde{\rho}_0(u+\theta_j)\, \tilde{S}_{\langle
\bar{1}'2\rangle}\hat{T}_{\langle \bar{1}'2\rangle }^{
+}(u+\eta+i\pi)\tilde{S}_{\langle
\bar{1}'2\rangle}^{-1}.\label{opr-1012}
 \eea

Next, we should combine the fusion relations (\ref{opr-1})-(\ref{opr-11}) of monodromy  matrices and the relations (\ref{op22r-1})-(\ref{opr-1012}) of reflecting monodromy matrices,
which can be achieved by the fusion of reflection matrices and that of dual ones. All the necessary fusion identities of reflection matrices have been deduced in appendices A and B   as Eqs.(\ref{8005-1})-(\ref{0805-2}), (\ref{0805-3})-(\ref{0805-31}).
According to the definitions of fused transfer matrices and through the direct calculation, we have
\bea &&
t(u)t(u+\Delta)=[\rho_1(2u+\Delta-4\eta)]^{-1}tr_{12}\{\bar{K}^{v}_{2}(u+\Delta)M_2^{-1}R_{12}^{vv}(-2u+8\eta-\Delta) \no\\
&&\quad\times M_2\bar{K}_1^{ v}(u)T_1 (u)
T_{2}(u+\Delta)K^{ v}_1(u)R_{21}^{vv}(2u+\Delta)K^{
v}_{2}(u+\Delta)\hat{T}_1
(u)\hat{T}_{2}(u+\Delta)\},\label{tt-1}  \\
&&t(u)t_{\pm}(u+\Delta)=[\rho_s(2u+\Delta-4\eta-i\pi)]^{-1}tr_{12}\{\bar{K}^{s_{\pm}}_{2}(u+\Delta)
\bar{M}_2^{-1} \no\\
&&\quad\times R_{12}^{vs_{\pm}}(-2u+8\eta+2i\pi-\Delta) \bar{M}_2\bar{K}_1^{ v}(u)T_1 (u)
T_{2}^{\pm}(u+\Delta)K^{ v}_1(u)\no \\
&& \quad\times
R_{21}^{s_{\pm}v}(2u+\Delta)K^{
s_{\pm}}_{2}(u+\Delta)\hat{T}_1
(u)\hat{T}_{2}^{{\pm}}(u+\Delta)\}, \label{tt-2} \\
 &&
t_+(u)t_-(u+\Delta)=[\rho_{ss}(2u-4\eta-i\pi+\Delta)]^{-1}tr_{12}\{\bar{K}^{s_{-}}_{2}
(u+\Delta)\bar{M}_2^{-1} \no\\
&&\quad\times R_{12}^{s_+s_{-}}(-2u+8\eta+2i\pi-\Delta)\bar{M}_2\bar{K}_1^{ s_+}(u)T_1^+ (u)
T_{2}^{-}(u+\Delta)K^{ s_+}_1(u)\no \\
&& \quad\times R_{21}^{s_{-}s_+}(2u+\Delta)K^{
s_{-}}_{2}(u+\Delta)\hat{T}_1^+
(u)\hat{T}_{2}^{-}(u+\Delta)\},\label{0804-12} \eea
where $\Delta$ is the shift of spectral parameter.
We find that if $\Delta$ is chosen as some suitable values such as $4\eta$, $2\eta+i\pi$, $3\eta+i\pi$ and $0$ in Eqs.(\ref{tt-1})-(\ref{0804-12}),
we can connect the fusions of monodromy matrices and that of the reflecting ones, and obtain the correct fusion relations among the fused transfer matrices $t(u)$ and $t_{\pm}(u)$.

Substituting Eqs.(\ref{opr-1})-(\ref{opr-11}),
(\ref{8005-1})-(\ref{0805-2}), (\ref{0805-3})-(\ref{0805-31}),
(\ref{op22r-1})-(\ref{opr-1012}) into Eq.(\ref{tt-1}) and considering $\{u=\pm\theta_j, \Delta=4\eta\}$,
into Eq.(\ref{tt-1}) with $\{u=\pm\theta_j, \Delta=2\eta+i\pi\}$ and using the result of Eq.(\ref{0804-12}) with the shift $\Delta=0$,
into Eq.(\ref{tt-2}) with $\{u=\pm\theta_j, \Delta=3\eta+i\pi\}$, we arrive at \bea &&
\hspace{-0.8cm}t(\pm\theta_j)t(\pm\theta_j+4\eta)=\cosh^2(\pm\theta_j)\frac{\sinh(\pm\theta_j-4\eta)
\sinh(\pm\theta_j+4\eta) }
{\sinh(\pm\theta_j-\eta)\sinh(\pm\theta_j+\eta)}
\no\\
&&\hspace{-0.8cm}\quad \times\frac{ \sinh(\pm 2\theta_j-6\eta)\sinh(\pm
2\theta_j+6\eta)}
{\sinh(\pm2\theta_j-4\eta)\sinh(\pm2\theta_j+4\eta)}\cosh(\pm\theta_j-\eta)\cosh(\pm\theta_j+\eta)\no\\
&&\hspace{-0.8cm}\quad\times \prod_{l=1}^N
a_1(\pm\theta_j-\theta_l)e_1(\pm\theta_j-\theta_l+4\eta)
a_1(\pm\theta_j+\theta_l)e_1(\pm\theta_j+\theta_l+4\eta)\times {\rm id},\no  \\
&&\hspace{-0.8cm}
t(\pm\theta_j)t(\pm\theta_j+2\eta+i\pi)=\frac{\sinh(\pm\theta_j+2\eta)\sinh(\pm\theta_j-4\eta)\cosh(\pm\theta_j+\eta)\cosh(\pm\theta_j-3\eta)}
{\sinh(\pm\theta_j+\eta)\sinh(\pm\theta_j-3\eta)\cosh(\pm\theta_j)\cosh(\pm\theta_j-2\eta)}\no\\
&&\hspace{-0.8cm}\quad\times\cosh^4 2\eta\prod_{l=1}^N
\tilde{\rho}_0(\pm\theta_j-\theta_l)\tilde{\rho}_0(\pm\theta_j+\theta_l)
 t_+(\pm\theta_j+\eta+i\pi)\,t_{-}(\pm\theta_j+\eta+i\pi),\no \\
 &&\hspace{-0.8cm}t(\pm\theta_j)t_{\pm}(\pm\theta_j+3\eta+i\pi)=4\frac{\cosh(\pm\theta_j)\cosh(\pm\theta_j+\eta)
\sinh(\pm\theta_j+3\eta)\sinh(\pm\theta_j-4\eta)}
{\sinh(\pm2\theta_j+2\eta)\sinh(\pm2\theta_j-4\eta)}\no\\
&&\hspace{-0.8cm}\quad\times\cosh(\pm\theta_j+\eta)\cosh(\pm\theta_j-3\eta)\prod_{l=1}^N
\tilde{\rho}_0(\pm\theta_j-\theta_l)\tilde{\rho}_0(\pm\theta_j+\theta_l)
 t_{\mp}(\pm\theta_j+\eta+i\pi),\label{Op-Product-Periodic-41} \eea
where $j=1,\cdots, N$.

The values of transfer matrices $t(u)$ and
$t_{\pm}(u)$ at the point of $u=0$ can be calculated directly \bea
t(0)=\frac{\sinh6\eta}{\sinh \eta}\prod_{j=1}^N\rho_1
(-\theta_j)\times {\rm id}, \quad
t_\pm(0)=-4\cosh^2\eta\prod_{j=1}^N\rho_s (-\theta_j)\times {\rm
id}. \label{2Op} \eea In the derivation, we have used the
relations \bea tr[\bar{K}^v(0)]K^v(0)=\frac{\sinh 6\eta}{\sinh
\eta}\times{\rm id},\quad
tr[\bar{K}^{s_{\pm}}(0)]K^{s_{\pm}}(0)=-4\cosh^2\eta\times{\rm
id}.\no\eea The asymptotic behaviors of $t(u)$ and $t_{\pm}(u)$  read
\bea t(u)|_{u\rightarrow
\pm\infty}=Q^v_{\pm}e^{\pm(4N+2)u}+\cdots, \quad
t_{\pm}(u)|_{u\rightarrow
\pm\infty}=Q^{s_{\pm}}_{\pm}e^{\pm(2N+2)u}+\cdots, \label{1Op}
\eea where $Q^v_{\pm}$ and $Q^{s_\pm}_{\pm}$ are the conserved
quantities with the definitions \bea &&Q^v_+=\frac
14\Big\{\frac{c'}{c}\Big(e^{2\eta}[T^v_+]^5_5[\hat{T}^v_+]^1_1-e^{-2\eta}[T^v_+]^6_5[\hat{T}^v_+]^1_2
+e^{-2\eta}[T^v_+]^6_6[\hat{T}^v_+]^2_2\Big)\no\\
&&\hspace{10mm}+e^{-4\eta}\Big([T^v_+]^3_3[\hat{T}^v_+]^3_3+
[T^v_+]^4_3[\hat{T}^v_+]^3_4
+[T^v_+]^4_4[\hat{T}^v_+]^4_4\Big)\no\\
&&\hspace{10mm}+\frac{c}{c'}\Big(e^{-6\eta}[T^v_+]^1_1[\hat{T}^v_+]^5_5
-e^{-10\eta}[T^v_+]^2_1[\hat{T}^v_+]^5_6
+e^{-10\eta}[T^v_+]^2_2[\hat{T}^v_+]^6_6\Big)\Big\},\no\\
&&Q^v_-=\frac
14\Big\{\frac{c'}{c}\Big(e^{10\eta}[T^v_-]^5_5[\hat{T}^v_-]^1_1-e^{10\eta}[T^v_-]^5_6[\hat{T}^v_-]^2_1
+e^{6\eta}[T^v_-]^6_6[\hat{T}^v_-]^2_2\Big)\no\\
&&\hspace{10mm}+e^{4\eta}\Big([T^v_-]^3_3[\hat{T}^v_-]^3_3+
[T^v_-]^3_4[\hat{T}^v_-]^4_3
+[T^v_-]^4_4[\hat{T}^v_-]^4_4\Big)\no\\
&&\hspace{10mm}+\frac{c}{c'}\Big(e^{2\eta}[T^v_-]^1_1[\hat{T}^v_-]^5_5
-e^{2\eta}[T^v_-]^1_2[\hat{T}^v_-]^6_5
+e^{-2\eta}[T^v_-]^2_2[\hat{T}^v_-]^6_6\Big)\Big\}, \no\\
&&Q^{s_{\pm}}_{\pm}=-\frac {1}{4e^{\pm 4\eta}\cosh^2
2\eta}\Big\{\frac{c'}{c}e^{4\eta}[T^{s_{\pm}}_\pm]^4_4[\hat{T}^{s_{\pm}}_\pm]^1_1
+\frac{c}{c'}e^{-4\eta}[T^{s_{\pm}}_\pm]^1_1[\hat{T}^{s_{\pm}}_\pm]^4_4\no\\
&&\hspace{10mm}+e^{2\eta}[T^{s_{\pm}}_\pm]^2_2[\hat{T}^{s_{\pm}}_\pm]^2_2+
e^{\mp 2\eta}[T^{s_{\pm}}_\pm]^3_2[\hat{T}^{s_{\pm}}_\pm]^2_3+
e^{-2\eta}[T^{s_{\pm}}_\pm]^3_3[\hat{T}^{s_{\pm}}_\pm]^3_3\Big\}.
 \eea
Here $[{T}_{\pm}^v]^\alpha_\beta$ and
$[{T}_{\pm}^{s_\pm}]^\alpha_\beta$ are given by (\ref{tjj-1}),
$[\hat{T}_{\pm}^v]^\alpha_\beta$ and
$[\hat{T}_{\pm}^{s_\pm}]^\alpha_\beta$ are the operators acting on
the quantum space ${\rm\bf V}\otimes {\rm\bf
V}\otimes\cdots\otimes{\rm\bf V}$ with the explicit expressions \bea &&[\hat
T_{\pm}^{v}]^{\alpha}_{\beta}=\sum_{\{\delta_i\}=1,\{\gamma_i\}=1}^6
[R^{vv(\pm)}_{10}]^{\gamma_1\alpha }_{ \delta_1\alpha_1}
[R^{vv(\pm)}_{20}]^ {\gamma_2 \alpha_1}_{ \delta_2\alpha_2} \cdots
[R^{vv(\pm)}_{N0}]^{\gamma_N\alpha_{N-1} }_{\delta_N \beta},\no\\[6pt]
&&[\hat T_{\pm}^{s_\pm}]^{\alpha}_{\beta}=
\sum_{\{\delta_i\}=1,\{\gamma_i\}=1}^6[R^{vs_\pm
(\pm)}_{1\tilde{0}'}]^{\gamma_1\alpha }_{ \delta_1\alpha_1}
[R^{vs_\pm (\pm)}_{2\tilde{0}'}]^{\gamma_2\alpha_1
}_{\delta_2\alpha_2 } \cdots [R^{vs_\pm
(\pm)}_{N\tilde{0}'}]^{\gamma_N\alpha_{N-1} }_{\delta_N \beta},
\label{tjj-3} \eea $R^{vv(\pm)}_{j0}$ and
$R^{vs_{\pm}(\pm)}_{j\tilde{0}'}$ are the leading terms of $e^{\mp
2u}R^{vv}_{j0}(u)$ and $e^{\mp u}R^{vs_{\pm}}_{j\tilde{0}'}(u)$
with $u\rightarrow\pm\infty$, respectively, and the repeated
indicators should be summarized. The detailed calculation shows
that the eigenvalues of conserved quantities $Q^v_{\pm}$ and
$Q^{s_\pm}_{\pm}$ can be characterized by the quantum number $m$
as \bea && \Lambda_{Q^v_{\pm}}= \frac
12\Big[\big(\frac{c}{c'}e^{-4\eta}+\frac{c'}{c}e^{4\eta}\big)\cosh
(2m\eta)+1\Big]e^{\pm(-8N\eta-4\eta)}, \no\\[8pt]
&& \Lambda_{Q^{s_\pm}_{\pm}}=-\frac {1}{4\cosh^2
2\eta}\Big[\frac{c}{c'}e^{-4\eta}+\frac{c'}{c}e^{4\eta}+2\cosh
(2m\eta)\Big]e^{\pm(-4N\eta-4\eta)}, \eea where $m\in[1,N+1]$. Then
we obtain the asymptotic behaviors of $t(u)$ and $t_{\pm}(u)$ as
\bea &&\hspace{-0.8cm}t(u)|_{u\rightarrow \pm\infty}= \frac
12\Big[\big(\frac{c}{c'}e^{-4\eta}+\frac{c'}{c}e^{4\eta}\big)\cosh
(2m\eta)+1\Big]e^{\pm(4Nu+2u-8N\eta-4\eta)}+\cdots,\no\\[8pt]
&&\hspace{-0.8cm}t_\pm(u)|_{u\rightarrow \pm\infty}=-\frac
{\frac{c}{c'}e^{-4\eta}+\frac{c'}{c}e^{4\eta}+2\cosh
(2m\eta)}{4\cosh^2
2\eta}e^{\pm(2Nu+2u-4N\eta-4\eta)}+\cdots.\label{14Op} \eea

Acting the fused transfer matrices $t(u)$ and $t_{\pm}(u)$ on a common eigenstate, we obtain the eigenvalues.
Denote the eigenvalues of $t(u)$ and
$t_{\pm}(u)$ as $\Lambda(u)$ and $\Lambda_{\pm}(u)$, respectively.
From the operators product identities (\ref{Op-Product-Periodic-41}), we obtain the functional
relations among the eigenvalues $\Lambda(u)$ and $\Lambda_{\pm}(u)$ as \bea &&
\hspace{-0.6cm}\Lambda(\pm\theta_j)\Lambda(\pm\theta_j+4\eta)=\cosh^2(\pm\theta_j)\frac{\sinh(\pm\theta_j-4\eta)
\sinh(\pm\theta_j+4\eta) }
{\sinh(\pm\theta_j-\eta)\sinh(\pm\theta_j+\eta)}
\no\\[6pt]
&&\hspace{-0.6cm}\quad \times\frac{ \sinh(\pm 2\theta_j-6\eta)\sinh(\pm
2\theta_j+6\eta)}
{\sinh(\pm2\theta_j-4\eta)\sinh(\pm2\theta_j+4\eta)}\cosh(\pm\theta_j-\eta)\cosh(\pm\theta_j+\eta)\no\\[6pt]
&&\hspace{-0.6cm}\quad\times \prod_{l=1}^N
a_1(\pm\theta_j-\theta_l)e_1(\pm\theta_j-\theta_l+4\eta)
a_1(\pm\theta_j+\theta_l)e_1(\pm\theta_j+\theta_l+4\eta),\no  \\[8pt]
&&\hspace{-0.6cm}
\Lambda(\pm\theta_j)\Lambda(\pm\theta_j+2\eta+i\pi)=\frac{\sinh(\pm\theta_j+2\eta)\sinh(\pm\theta_j-4\eta)\cosh(\pm\theta_j+\eta)\cosh(\pm\theta_j-3\eta)}
{\sinh(\pm\theta_j+\eta)\sinh(\pm\theta_j-3\eta)\cosh(\pm\theta_j)\cosh(\pm\theta_j-2\eta)}\no\\[6pt]
&&\hspace{-0.6cm}\quad\times\cosh^4 2\eta\prod_{l=1}^N
\tilde{\rho}_0(\pm\theta_j-\theta_l)\tilde{\rho}_0(\pm\theta_j+\theta_l)
 \Lambda_+(\pm\theta_j+\eta+i\pi)\Lambda_{-}(\pm\theta_j+\eta+i\pi),\no \\[8pt]
 &&\hspace{-0.6cm}\Lambda(\pm\theta_j)\,\Lambda_{\pm}(\pm\theta_j+3\eta+i\pi)=4\frac{\cosh(\pm\theta_j)\cosh(\pm\theta_j+\eta)
\sinh(\pm\theta_j+3\eta)\sinh(\pm\theta_j-4\eta)}
{\sinh(\pm2\theta_j+2\eta)\sinh(\pm2\theta_j-4\eta)}\no\\[6pt]
&&\hspace{-0.6cm}\quad\times\cosh(\pm\theta_j+\eta)\cosh(\pm\theta_j-3\eta)\prod_{l=1}^N
\tilde{\rho}_0(\pm\theta_j-\theta_l)\tilde{\rho}_0(\pm\theta_j+\theta_l)
 \Lambda_{\mp}(\pm\theta_j+\eta+i\pi), \label{Op-e4} \eea
where $j=1,\cdots, N$. According to Eq.(\ref{2Op}), the values of $\Lambda(u)$ and
$\Lambda_{\pm}(u)$ at the point of $u=0$ are \bea
\Lambda(0)=\frac{\sinh 6\eta}{\sinh\eta}\prod_{j=1}^N\rho_1
(-\theta_j), \quad
\Lambda_+(0)=\Lambda_-(0)=-4\cosh^2\eta\prod_{j=1}^N\rho_s
(-\theta_j).\label{3Op} \eea Eq.(\ref{1Op}) gives the asymptotic
behaviors of $\Lambda(u)$ and $\Lambda_{\pm}(u)$ \bea
&&\Lambda(u)|_{u\rightarrow \pm\infty}= \frac
12\big[\big(\frac{c}{c'}e^{-4\eta}+\frac{c'}{c}e^{4\eta}\big)\cosh
(2m\eta)+1\big]e^{\pm(4Nu+2u-8N\eta-4\eta)}+\cdots,\no\\[8pt]
&&\Lambda_\pm(u)|_{u\rightarrow \pm\infty}=- \frac
{\frac{c}{c'}e^{-4\eta}+\frac{c'}{c}e^{4\eta}+2\cosh
2m\eta}{4\cosh^2
2\eta}e^{\pm(2Nu+2u-4N\eta-4\eta)}+\cdots.\label{4Op} \eea From
the definition, we know that the eigenvalues $\Lambda(u)$ and
$\Lambda_{\pm}(u)$ are the polynomials of $e^u$ with degrees
$4N+2$ and $2N+2$, respectively. Therefore, the $8N+9$ constraints
(\ref{Op-e4})-(\ref{4Op}) can completely determine the values of
$\Lambda(u)$ and $\Lambda_{\pm}(u)$.

\subsection{Inhomogeneous T-Q relations}

The functional relations (\ref{Op-e4}), the values (\ref{3Op}) of eigenvalues at $u=0$  and the asymptotic
behaviors (\ref{4Op}) allow us to determinate the eigenvalues of the corresponding transfer matrices, which are given in terms of some inhomogeneous $T-Q$ relations.  The explicit expressions of the
eigenvalues $\Lambda(u)$ and
$\Lambda_{\pm}(u)$ read
\bea
&&\Lambda(u)=2\frac{\cosh u\sinh(u-4\eta)\sinh(u-3\eta)}{\sinh(u-\eta)\sinh(2u-4\eta)}\prod_{j=1}^N
a_1(u-\theta_j)a_1(u+\theta_j)\no\\
&&\qquad \times \cosh(u+\eta)\cosh(u-3\eta)\,\frac{Q^{(1)}(u+2\eta)}{Q^{(1)}(u)}\no\\[4pt]
&&\qquad +2\frac{\sinh u \cosh(u-4\eta)\sinh(u-\eta)}{\sinh(u-3\eta)\sinh(2u-4\eta)}\prod_{j=1}^N
e_1(u-\theta_j)e_1(u+\theta_j)\no\\
&&\qquad \times \cosh(u-\eta) \cosh(u-5\eta)
\frac{Q^{(1)}(u-4\eta)}{Q^{(1)}(u-2\eta)}\no\\[4pt]
&&\qquad+\frac{\sinh u \sinh(u-4\eta)}{\sinh(u-\eta)\sinh(u-3\eta)
}\prod_{j=1}^Nb_1(u-\theta_j)b_1(u+\theta_j)\frac{1}{Q^{(1)}(u)Q^{(1)}(u-2\eta)}\no\\
&&\qquad \times\Big[\frac{\sinh(2u-6\eta)}{\sinh(2u-4\eta)
}Q^{(1)}(u-2\eta)\frac{Q^{(2)}(u+2\eta)}
{Q^{(2)}(u)}\cosh u\no\\
&&\qquad +\frac{\sinh(2u-2\eta)}{\sinh(2u-4\eta)
}Q^{(1)}(u)\frac{Q^{(2)}(u-2\eta)}
{Q^{(2)}(u)}\cosh(u-4\eta)\Big]\no\\
&&\qquad \times\Big[\frac{\sinh(2u-6\eta)}{\sinh(2u-4\eta)
}Q^{(1)}(u-2\eta)\frac{Q^{(3)}(u+2\eta)}
{Q^{(3)}(u)}\cosh u\no\\
&&\qquad +\frac{\sinh(2u-2\eta)}{\sinh(2u-4\eta)
}Q^{(1)}(u)\frac{Q^{(3)}(u-2\eta)}
{Q^{(3)}(u)}\cosh(u-4\eta)\Big]\no\\
&&\qquad+
4^Nh\prod_{j=1}^Na_1(u-\theta_j)a_1(u+\theta_j)\sinh(u-\theta_j)\sinh(u+\theta_j)
\no\\
&&\qquad\times\frac{\sinh u\sinh(u-4\eta)}{\sinh(2u-4\eta)}\Big[
\cosh^2 u\sinh(2u-6\eta)
 \frac{Q^{(2)}(u+2\eta)Q^{(3)}(u+2\eta)}{Q^{(1)}(u)}\no\\
&&\qquad + \cosh^2(u-4\eta)\sinh(2u-2\eta)
\frac{Q^{(2)}(u-2\eta)Q^{(3)}(u-2\eta)}{Q^{(1)}(u-2\eta)}\Big],\label{tannn1} \\
&&\Lambda_+(u)=-\frac{1}{\cosh^22\eta}\Big\{\prod_{j=1}^N
a_2(u-\theta_j)a_2(u+\theta_j)\frac{\sinh(u-4\eta)}{\sinh(u-\eta)}\no\\
&&\qquad \times
\left[\cosh(u+\eta)\cosh(u-2\eta)\frac{Q^{(2)}(u+3\eta)}{Q^{(2)}(u+\eta)}\right.\no\\
&&\qquad
\left. +\frac{\sinh u}{\sinh(u-2\eta)}\cosh u \cosh(u-3\eta)
\frac{Q^{(1)}(u+\eta)Q^{(2)}(u-\eta)}{Q^{(1)}(u-\eta)Q^{(2)}(u+\eta)}\right]\no\\
&&\qquad +\prod_{j=1}^N
b_2(u-\theta_j)b_2(u+\theta_j)\frac{\sinh u}{\sinh(u-3\eta)}\no\\
&&\qquad \times\left[\cosh(u-5\eta)\cosh(u-2\eta)
\frac{Q^{(3)}(u-3\eta)}{Q^{(3)}(u-\eta)}\right.\no\\
&&\qquad
\left.+\frac{\sinh(u-4\eta)}{\sinh(u-2\eta)}\cosh(u-\eta)\cosh(u-4\eta)
\frac{Q^{(1)}(u-3\eta)Q^{(3)}(u+\eta)}{Q^{(1)}(u-\eta)Q^{(3)}(u-\eta)}\right]\no\\
&&\qquad +h\,
\sinh u\sinh(u-4\eta)\cosh(u-\eta)\cosh(u-3\eta)\no\\
&&\qquad \times\prod_{j=1}^Na_2(u-\theta_j)a_2(u+\theta_j)
b_2(u-\theta_j)b_2(u+\theta_j)\frac{Q^{(2)}(u-\eta)Q^{(3)}(u+\eta)}{Q^{(1)}(u-\eta)}\Big\}, \label{tannn2}\\
&&\Lambda_-(u)=-\frac{1}{\cosh^22\eta}\Big\{\prod_{j=1}^N
a_2(u-\theta_j)a_2(u+\theta_j)\frac{\sinh(u-4\eta)}{\sinh(u-\eta)}\no\\
&&\qquad \times
\left[\cosh(u+\eta)\cosh(u-2\eta)\frac{Q^{(3)}(u+3\eta)}{Q^{(3)}(u+\eta)}\right.\no\\
&&\qquad
\left.+\frac{\sinh u }{\sinh(u-2\eta)}\cosh u\cosh(u-3\eta)
\frac{Q^{(1)}(u+\eta)Q^{(3)}(u-\eta)}{Q^{(1)}(u-\eta)Q^{(3)}(u+\eta)}\right]\no\\
&&\qquad +\prod_{j=1}^N
b_2(u-\theta_j)b_2(u+\theta_j)\frac{\sinh u}{\sinh(u-3\eta)}\no\\
&&\qquad \times\left[\cosh(u-5\eta)\cosh(u-2\eta)
\frac{Q^{(2)}(u-3\eta)}{Q^{(2)}(u-\eta)}\right.\no\\
&&\qquad
\left.+\frac{\sinh(u-4\eta)}{\sinh(u-2\eta)}\cosh(u-\eta)\cosh(u-4\eta)
\frac{Q^{(1)}(u-3\eta)Q^{(2)}(u+\eta)}{Q^{(1)}(u-\eta)Q^{(2)}(u-\eta)}\right]\no\\
&&\qquad +h\,
\sinh u\sinh(u-4\eta)\cosh(u-\eta)\cosh(u-3\eta)\no\\
&&\qquad \times\prod_{j=1}^Na_2(u-\theta_j)a_2(u+\theta_j)
b_2(u-\theta_j)b_2(u+\theta_j)\frac{Q^{(3)}(u-\eta)Q^{(2)}(u+\eta)}{Q^{(1)}(u-\eta)}\Big\},\label{tannn3} \eea
where $h$ is an undetermined parameter, the $Q$-functions are defined by
\bea
&&Q^{(1)}(u)=\prod_{k=1}^{L_1}\sinh(u-\mu_k^{(1)}-\eta)\sinh(u+\mu_k^{(1)}-\eta),\no\\
&&Q^{(2)}(u)=Q^{(3)}(u-i\pi)=\prod_{l=1}^{L_2}\sinh\frac12(u-\mu_l^{(2)}-2\eta)\sinh\frac12(u+\mu_l^{(2)}-2\eta),\eea
$L_1$ is the number of Bethe roots $\{\mu_k^{(1)}\}$ and $L_2$ is the number of Bethe roots $\{\mu_k^{(2)}\}$.

The regularities of eigenvalues $\Lambda(u)$ and
$\Lambda_{\pm}(u)$ require that the Bethe roots $\{\mu^{(1)}_k\}$ and $\{\mu^{(2)}_l\}$
satisfy the BAEs \bea
&&\hspace{-10mm}\frac{2\sinh(2\mu_k^{(1)}-2\eta)\cosh(\mu_k^{(1)}+2\eta)}{4^N\prod_{j=1}^N\sinh(\mu_k^{(1)}+\eta-\theta_j)\sinh(\mu_k^{(1)}+\eta+\theta_j)}
\frac{Q^{(1)}(\mu_k^{(1)}+3\eta)}{Q^{(2)}(\mu_k^{(1)}+3\eta)Q^{(3)}(\mu_k^{(1)}+3\eta)}\no\\[6pt]
&&\hspace{-10mm}\quad\quad\quad\quad
+\frac{2\sinh(2\mu_k^{(1)}+2\eta)\cosh(\mu_k^{(1)}-2\eta)}{4^N\prod_{j=1}^N\sinh(\mu_k^{(1)}-\eta-\theta_j)\sinh(\mu_k^{(1)}-\eta+\theta_j)}
\frac{Q^{(1)}(\mu_k^{(1)}-\eta)}{Q^{(2)}(\mu_k^{(1)}+\eta)Q^{(3)}(\mu_k^{(1)}+\eta)}\no\\[8pt]
&&\hspace{-10mm}\quad\quad
=-h\,\ \sinh \mu_k^{(1)} \sinh(2\mu_k^{(1)}-2\eta)\sinh(2\mu_k^{(1)}+2\eta), \quad k=1,\cdots, L_1, \label{BAEs-2231} \\[6pt]
&&\hspace{-10mm}\frac{Q^{(1)}(\mu_l^{(2)})Q^{(2)}(\mu_l^{(2)}+4\eta)}{Q^{(1)}(\mu_l^{(2)}+2\eta)Q^{(2)}(\mu_l^{(2)})}
=-\frac{\sinh(2\mu_l^{(2)}+2\eta)\cosh(\mu_l^{(2)}-2\eta)}
{\sinh(2\mu_l^{(2)}-2\eta)\cosh(\mu_l^{(2)}+2\eta)}, \quad
l=1,\cdots, L_2,  \label{BAEs-223} \eea where the numbers of Bethe
roots should satisfy the constraint \bea L_1=L_2+N+1, \eea and the
parameter $h$ is \bea
h=2^{2L_2-2N}\Big\{\frac{c}{c'}e^{-4\eta}+\frac{c'}{c}e^{4\eta}-2\cosh[2(L_1+1)\eta]\Big\}.\eea

Some remarks are in order. The BAEs (\ref{BAEs-2231}) are inhomogeneous while the BAEs (\ref{BAEs-223}) are homogeneous.
This is because  the reflection matrices (\ref{K-matrix-VV}) and (\ref{ksk111}) can be
divided into the direct summation of a $4\times 4$ non-diagonal and a $2\times 2$ diagonal submatrices.
The boundary reflection in the non-diagonal subspace breaks the $U(1)$ symmetry of the system.
While in the  diagonal subspace, there exists a conserved charge, which leads to the homogeneous BAEs (\ref{BAEs-223}).
From the asymptotic behavior of the eigenvalues, we obtain that the quantum number $m$ of conserved quantities $Q^v_{\pm}$ and $Q^{s_\pm}_{\pm}$ is related with the number of Bethe roots
$\{\mu^{(2)}_l\}$ as $m=L_2-N$, which is consistent with the conclusion that BAEs (\ref{BAEs-223}) are homogeneous.
We shall also note that the BAEs obtained from the regularities of
$\Lambda(u)$ are the same as those obtained from the regularities of
$\Lambda_{\pm}(u)$. The function $Q^{(l)}(u)$ has two sets of zero roots, i.e., $\{\mu_k^{(l)}+l\eta|k=1,\cdots, L_l\}$ and $\{-\mu_k^{(l)}+l\eta|k=1,\cdots, L_l\}$, where $l=1, 2$.
The BAEs obtained from these two sets of zero roots are also the same. It is
easy to check that $\Lambda(u)$ (\ref{tannn1}) and $\Lambda_{\pm}(u)$ (\ref{tannn2})-(\ref{tannn3}) satisfy
the functional relations (\ref{Op-e4}), the values at the special
points (\ref{3Op}) and the asymptotic behaviors (\ref{4Op}).
Therefore, we conclude that the inhomogeneous $T-Q$ relations (\ref{tannn1})-(\ref{tannn3})
give the eigenvalues of the transfer matrices $t(u)$ and
$t_{\pm}(u)$. All the eigenvalues (\ref{tannn1})-(\ref{tannn3}) and BAEs (\ref{BAEs-2231})-(\ref{BAEs-223})
have the well-defined homogeneous limit.

The energy spectrum of the Hamiltonian (\ref{hh}) can be obtained by $\Lambda(u)$ as
\begin{eqnarray}
E= \frac{1}{2}\frac{\partial \ln \Lambda(u)}{\partial
u}|_{u=0,\{\theta_j\}=0}.
\end{eqnarray}

\section{Discussion}

In this paper, we have studied the quantum integrable model associated with the twisted  $D^{(2)}_3$ Lie algebra by generalizing the nested off-diagonal Bethe ansatz.
We obtain the exact solutions of the system with either periodic or non-diagonal open boundary conditions.
With the help of fusion, we obtain the closed recursive operator product identities among the fused transfer matrices.
Based on them and the asymptotic behaviors as
well as the special values at certain points, we obtain the eigen-spectrum and Bethe ansatz equations.
For the periodic case, the eigenvalues of the transfer matrices are described by the homogeneous $T-Q$ relations.
While for the open boundary case, the eigenvalues are characterized by the inhomogeneous $T-Q$ relations due to the off-diagonal $K$-matrices (\ref{K-matrix-VV})-(\ref{ksk111}).
The method  can be generalized to the models with  other
high rank twisted algebras.

%%%%%%%%%%%%%%%%%%%%%%%%%%%%%%%%%%%%%%%%%%%%%%%%%%%%%%%%%%%%%%%
%                                                             %
%  Acknowledgments                                            %
%                                                             %
%%%%%%%%%%%%%%%%%%%%%%%%%%%%%%%%%%%%%%%%%%%%%%%%%%%%%%%%%%%%%%%
\section*{Acknowledgments}

The financial supports from the National
Natural Science Foundation of China (Grant Nos. 12074410, 12047502, 12075177, 11934015,
11975183, 12105221,
91536115 and 11805152), Major Basic Research Program of Natural Science of
Shaanxi Province (Grant Nos. 2021JCW-19, 2017ZDJC-32), Australian
Research Council (Grant No. DP 190101529), Strategic
Priority Research Program of the Chinese Academy of Sciences (Grant No. XDB33000000), Shaanxi Province Key Laboratory
of Quantum Information and Quantum Optoelectronic Devices, Xi'an
Jiaotong University, and Double First-Class University Construction Project of Northwest
University are gratefully acknowledged.

%%%%%%%%%%%%%%%%%%%%%%%%%%%%%%%%%%%%%%%%%%%%%%%%%%%%%%%%%%%%%%%%
%                                                             %
%  Appendix A                                                 %
%                                                             %
%                                                             %
%                                                             %
%%%%%%%%%%%%%%%%%%%%%%%%%%%%%%%%%%%%%%%%%%%%%%%%%%%%%%%%%%%%%%%

\section*{Appendix A: Fusion of the $R$-matrices }
\setcounter{equation}{0}
\renewcommand{\theequation}{A.\arabic{equation}}

\subsection*{A.1 Spinorial $R$-matrix}

In this appendix, we shall give the $R$-matrices $R^{s_{\pm}v}(u)$ which has the same quantum space as that of (\ref{Rv}) but  the spinorial representations of $D^{(2)}_3$ as their auxiliary spaces.
Namely,  we provide $R^{s_{+}v}(u)$ as
\begingroup
\renewcommand*{\arraystretch}{0.1}
\begin{equation}
    \begin{pmatrix}\setlength{\arraycolsep}{1.0pt}
    \begin{array}{cccccc|cccccc|cccccc|cccccc}
    r^+_1&&&&& &&&&&& &&&&&& &&&&&&  \\
    &r^+_1&&&& &&&&&& &&&&&& &&&&&& \\
    &&r^+_3&r^+_5&& &&r^+_{19}&&&& &r^+_{23}&&&&& &&&&&&  \\
    &&r^+_{6}&r^+_{4}&& &&r^+_{20}&&&& &r^+_{24}&&&&& &&&&&&  \\
    &&&&r^+_{2}& &&&r^+_{21}&r^+_{22}&& &&&&&& &r^+_{15}&&&&&  \\
    &&&&&r^+_{2} &&&&&& &&&-r^+_{21}&-r^+_{22}&& &&-r^+_{15}&&&& \\
   \hline &&&&& &r^+_{1}&&&&& &&&&&& &&&&&&  \\
    &&r^+_{25}&r^+_{26}&& &&r^+_{2}&&&& &r^+_{16}&&&&& &&&&&& \\
    &&&&r^+_{27}& &&&r^+_{7}&&& &&&&&& &r^+_{31}&&&&& \\
    &&&&r^+_{28}& &&&r^+_{14}&r^+_{8}&& &&&&&& &r^+_{32}&&&&&  \\
     &&&&& &&&&&r^+_{2}& &&&&&& &&&&&& \\
     &&&&& &&&&&&r^+_{1} &&&&&-r^+_{16}& &&&r^+_{33}&r^+_{34}&&  \\
   \hline  &&r^+_{29}&r^+_{30}&& &&r^+_{17}&&&& &r^+_{2}&&&&& &&&&&&  \\
    &&&&& &&&&&& &&r^+_{1}&&&& &&&&&&  \\
   &&&&&-r^+_{27} &&&&&& &&&r^+_{7}&&& &&-r^+_{31}&&&&  \\
   &&&&&-r^+_{28} &&&&&& &&&-r^+_{14}&r^+_{8}&& &&r^+_{32}&&&& \\
     &&&&& &&&&&&-r^+_{17} &&&&&r^+_{2}& &&&r^+_{35}&r^+_{36}&&  \\
      &&&&& &&&&&& &&&&&&r^+_{1} &&&&&&  \\
   \hline &&&&r^+_{18}& &&&r^+_{37}&r^+_{38}&& &&&&&& &r^+_{2}&&&&&  \\
    &&&&&-r^+_{18} &&&&&& &&&r^+_{37}&-r^+_{38}&& &&r^+_{2}&&&&  \\
    &&&&&  &&&&&&r^+_{39} &&&&&r^+_{11}& &&&r^+_{9}&-r^+_{5}&& \\
     &&&&& &&&&&&r^+_{40} &&&&&r^+_{12}& &&&r^+_{13}&r^+_{10}&& \\
     &&&&& &&&&&& &&&&&& &&&&&r^+_{1}&  \\
      &&&&& &&&&&& &&&&&& &&&&&&r^+_{1}  \\
\end{array}
    \end{pmatrix},\label{rsp}
\end{equation}\noindent
\endgroup
where \bea
&&\hspace{-0.5cm}r^+_{1}=2\sinh(u-3\eta),\
r^+_{2}=2\sinh(u-\eta), \ r^+_{3}=2\sinh(u-2\eta)+\frac{e^{\eta}\sinh 4\eta-2e^{2\eta}\sinh
\eta}{\cosh 2\eta}, \no\\[6pt]
&&\hspace{-0.5cm} r^+_{4}=2\sinh(u-2\eta)-\frac{e^{\eta}\sinh
4\eta-2e^{2\eta}\sinh \eta}{\cosh 2\eta},\ r^+_{5}=-2\sinh
\eta\tanh 2\eta\sqrt{\frac{\cosh 3\eta}{\cosh \eta}},\no\\[6pt]
&&\hspace{-0.5cm}
r^+_{6}=\frac{2\sinh^2\eta(1+e^{4\eta}\cosh 2\eta)}{\cosh
2\eta\sqrt{\cosh \eta\cosh 3\eta}},\ r^+_{7}=4\sinh\frac
12(u-3\eta)\cosh\frac 12(u-\eta),\no\\[6pt]
&&\hspace{-0.5cm} r^+_{8}=4\sinh\frac
12(u-\eta)\cosh\frac 12(u-3\eta),\
r^+_{9}=2\sinh(u-2\eta)+\frac{e^{-\eta}\sinh
4\eta-2e^{-2\eta}\sinh
\eta}{\cosh 2\eta},\no\\[6pt]
&&\hspace{-0.5cm}r^+_{10}=2\sinh(u-2\eta)-\frac{e^{-\eta}\sinh
4\eta-2e^{-2\eta}\sinh \eta}{\cosh
2\eta},\no\\[6pt]
&&\hspace{-0.5cm} r^+_{11}=-4e^{\frac{u}{2}-\eta}\sinh\eta\cosh\frac
12(u-\eta)\sqrt{\frac{\cosh 3\eta}{\cosh 2\eta}},\no\\[6pt]
&&\hspace{-0.5cm} r^+_{12}=-2e^{-\frac{\eta}{2}}\sinh\eta(2\cosh
2\eta-1-e^{u-\eta})\sqrt{\frac{\cosh \eta}{\cosh 2\eta}},\
r^+_{13}=-\frac{2\sinh^2\eta(1+e^{-4\eta}\cosh 2\eta)}{\cosh
2\eta\sqrt{\cosh \eta\cosh 3\eta}}, \no \\[6pt]
&&\hspace{-0.5cm}   r^+_{14}=-\frac{4\sinh
2\eta\sinh^2\eta}{\sqrt{\cosh \eta\cosh 3\eta}},\
r^+_{15}=2e^{-u}\sin 2\eta,\ r^+_{16}=-2e^{-u+\eta}\sin
2\eta, \ r^+_{17}=-2e^{u-\eta}\sin 2\eta, \no\\[6pt]
&&\hspace{-0.5cm}
r^+_{18}=2e^{u}\sin 2\eta,\
r^+_{19}=e^{-u+2\eta}r^+_{11}, \ r^+_{20}=-2e^{\frac{\eta}{2}}\sinh\eta(2\cosh
2\eta-1-e^{\eta-u})\sqrt{\frac{\cosh \eta}{\cosh 2\eta}},\no\\[6pt]
&&\hspace{-0.5cm}  r^+_{21}=\frac{e^{-u+\frac{3\eta}{2}}\sinh 2\eta}{\sqrt{{\cosh
2\eta}{\cosh 3\eta}}}(2\cosh
2\eta-1-e^{u-3\eta}),\no\\[6pt]
&&\hspace{-0.5cm}  r^+_{22}=-4e^{-\frac{u}{2}}\sinh\eta\cosh(\frac
12(u-3\eta))\sqrt{\frac{\cosh \eta}{\cosh 2\eta}}, \
r^+_{23}=-e^{-u}r^+_{11},\  r^+_{24}=-e^{-2\eta}r^+_{20},\no\\[6pt]
&&\hspace{-0.5cm} r^+_{25}=-\frac{e^{u-\frac{3\eta}{2}}\sinh 2\eta}{\sqrt{{\cosh
2\eta}{\cosh 3\eta}}}(2\cosh
2\eta-1+e^{\eta-u}),\ r^+_{26}=4e^{\frac{u}{2}-\eta}\sinh\eta\sinh\frac
12(u-\eta)\sqrt{\frac{\cosh \eta}{\cosh 2\eta}},\no\\[6pt]
&&\hspace{-0.5cm} r^+_{27}=4e^{\frac{u}{2}}\sinh\eta\sinh(\frac
12(u-3\eta))\sqrt{\frac{\cosh 3\eta}{\cosh 2\eta}},\ r^+_{28}=-\frac{e^{\frac{3\eta}{2}}\sinh 2\eta}{\sqrt{{\cosh
\eta}{\cosh 2\eta}}}(2\cosh
2\eta-1+e^{u-3\eta}),\no\\[6pt]
&&\hspace{-0.5cm} r^+_{29}=-e^{2\eta}r^+_{25},\  r^+_{30}=-e^{2\eta}r^+_{26},\  r^+_{31}=-e^{-u}r^+_{26},\no\\
&&\hspace{-0.5cm} r^+_{32}=-\frac{e^{-\frac{3\eta}{2}}\sinh 2\eta}{\sqrt{{\cosh
\eta}{\cosh 2\eta}}}(2\cosh
2\eta-1+e^{3\eta-u}),\no\\[6pt]
&&\hspace{-0.5cm} r^+_{33}=-\frac{e^{-u-\frac{\eta}{2}}\sinh 2\eta}{\sqrt{{\cosh
2\eta}{\cosh 3\eta}}}(2\cosh
2\eta-1+e^{u-\eta}),\ r^+_{34}=-e^{-u}r^+_{26},\  r^+_{35}=e^{2\eta}r^+_{33},\no\\[6pt]
&&\hspace{-0.5cm}  r^+_{36}=-e^{-u+2\eta}r^+_{26},\
 r^+_{37}=\frac{e^{u-\frac{3\eta}{2}}\sinh 2\eta}{\sqrt{{\cosh
2\eta}{\cosh 3\eta}}}(2\cosh
2\eta-1-e^{3\eta-u}),\no\\[6pt]
&&\hspace{-0.5cm} r^+_{38}=e^u r^+_{22},\  r^+_{39}=e^{2\eta}r^+_{11},\
r^+_{40}=e^{2\eta}r^+_{12}, \no
 \eea
and $R^{s_{-}v}(u)$ as
\begingroup
\renewcommand*{\arraystretch}{0.1}
\begin{equation}
    \begin{pmatrix}\setlength{\arraycolsep}{1.0pt}
    \begin{array}{cccccc|cccccc|cccccc|cccccc}
    r^-_1&&&&& &&&&&& &&&&&& &&&&&&  \\
    &r^-_1&&&& &&&&&& &&&&&& &&&&&& \\
    &&r^-_3&r^-_5&& &&r^-_{19}&&&& &r^-_{23}&&&&& &&&&&&  \\
    &&r^-_{6}&r^-_{4}&& &&r^-_{20}&&&& &r^-_{24}&&&&& &&&&&&  \\
    &&&&r^-_{2}& &&&r^-_{21}&r^-_{22}&& &&&&&& &r^-_{15}&&&&&  \\
    &&&&&r^-_{2} &&&&&& &&&r^-_{21}&r^-_{22}&& &&-r^-_{15}&&&& \\
   \hline &&&&& &r^-_{1}&&&&& &&&&&& &&&&&&  \\
    &&r^-_{25}&r^-_{26}&& &&r^-_{2}&&&& &r^-_{16}&&&&& &&&&&& \\
    &&&&r^-_{27}& &&&r^-_{7}&&& &&&&&& &r^-_{31}&&&&& \\
    &&&&r^-_{28}& &&&r^-_{14}&r^-_{8}&& &&&&&& &r^-_{32}&&&&&  \\
     &&&&& &&&&&r^-_{2}& &&&&&& &&&&&& \\
     &&&&& &&&&&&r^-_{1} &&&&&-r^-_{16}& &&&r^-_{33}&r^-_{34}&&  \\
   \hline  &&r^-_{29}&r^-_{30}&& &&r^-_{17}&&&& &r^-_{2}&&&&& &&&&&&  \\
    &&&&& &&&&&& &&r^-_{1}&&&& &&&&&&  \\
   &&&&&r^-_{27} &&&&&& &&&r^-_{7}&&& &&-r^-_{31}&&&&  \\
   &&&&&r^-_{28} &&&&&& &&&r^-_{14}&r^-_{8}&& &&-r^-_{32}&&&& \\
     &&&&& &&&&&&-r^-_{17} &&&&&r^-_{2}& &&&r^-_{35}&r^-_{36}&&  \\
      &&&&& &&&&&& &&&&&&r^-_{1} &&&&&&  \\
   \hline &&&&r^-_{18}& &&&r^-_{37}&r^+_{38}&& &&&&&& &r^-_{2}&&&&&  \\
    &&&&&-r^-_{18} &&&&&& &&&-r^-_{37}&-r^-_{38}&& &&r^-_{2}&&&&  \\
    &&&&&  &&&&&&r^-_{39} &&&&&r^-_{11}& &&&r^-_{9}&-r^-_{5}&& \\
     &&&&& &&&&&&r^-_{40} &&&&&r^-_{12}& &&&r^-_{13}&r^-_{10}&& \\
     &&&&& &&&&&& &&&&&& &&&&&r^-_{1}&  \\
      &&&&& &&&&&& &&&&&& &&&&&&r^-_{1}  \\
\end{array}
    \end{pmatrix},\label{rsn}
\end{equation}\noindent
\endgroup
with \bea &&\hspace{-0.5cm}r^-_{1}=r^+_{1},\ r^-_{2}=r^+_{2},\
r^-_{3}=r^+_{4},\ r^-_{4}=r^+_{3},\  r^-_{5}=-r^+_{5},\
 r^-_{6}=-r^+_{6},\
    r^-_{7}=r^+_{8},\no\\[6pt]
&&\hspace{-0.5cm} r^-_{8}=r^+_{7},
    \no \
r^-_{9}=r^+_{10},\ r^-_{10}=r^+_{9},\ r^-_{11}=\sqrt{\frac{\cosh
3\eta}{\cosh \eta}}r^+_{26},\  r^-_{12}=e^{u}\sqrt{\frac{\cosh
3\eta}{\cosh \eta}}r^+_{33},\no\\[6pt]
&&\hspace{-0.5cm} r^-_{13}=-r^+_{13},\   r^-_{14}=-r^+_{14}, \
r^-_{15}=r^+_{15},\ r^-_{16}=-r^+_{16},\
r^-_{17}=-r^+_{17},\ r^-_{18}=r^+_{18},\no\\[6pt]
&&\hspace{-0.5cm}  r^-_{19}=e^{-u+2\eta}r^-_{11},\
r^-_{20}=-e^{2\eta-u}\sqrt{\frac{\cosh 3\eta}{\cosh
\eta}}r^+_{25}, \  r^-_{21}=-e^{-u}\sqrt{\frac{\cosh \eta}{\cosh
3\eta}}r^+_{28},\no\\[6pt]
&&\hspace{-0.5cm}  r^-_{22}=e^{-u}\sqrt{\frac{\cosh \eta}{\cosh
3\eta}}r^+_{27},\
 r^-_{23}=e^{-u}r^-_{11},\  r^-_{24}=e^{2\eta}r^-_{20},\  r^-_{25}=e^{u-2\eta}\sqrt{\frac{\cosh
\eta}{\cosh 3\eta}}r^+_{20},\no\\[6pt]
&&\hspace{-0.5cm}  r^-_{26}=-\sqrt{\frac{\cosh \eta}{\cosh
3\eta}}r^+_{11},\  r^-_{27}=-e^{u}\sqrt{\frac{\cosh 3\eta}{\cosh
\eta}}r^+_{22},\   r^-_{28}=e^{u}\sqrt{\frac{\cosh 3\eta}{\cosh
\eta}}r^+_{21},\no\\[6pt]
&&\hspace{-0.5cm} r^-_{29}=e^{2\eta}r^-_{25},\
r^-_{30}=e^{2\eta}r^-_{26},\ r^-_{31}=e^{-u}r^-_{27},\
r^-_{32}=e^{-u}\sqrt{\frac{\cosh 3\eta}{\cosh \eta}}r^+_{37},\no\\[6pt]
&&\hspace{-0.5cm} r^-_{33}=e^{-u}\sqrt{\frac{\cosh \eta}{\cosh
3\eta}}r^+_{12},\ r^-_{34}=e^{-u}r^-_{26},\
r^-_{35}=-e^{2\eta}r^-_{33},\ r^-_{36}=-e^{-u+2\eta}r^-_{26}, \no\\[6pt]
&&\hspace{-0.5cm} r^-_{37}=-e^{u}\sqrt{\frac{\cosh \eta}{\cosh
3\eta}}r^+_{32},\  r^-_{38}=-e^u r^-_{22} ,\
r^-_{39}=-e^{2\eta}r^-_{11},\ r^-_{40}=-e^{2\eta}r^-_{12}.
 \eea

We remark that the fused $R$-matrices
$R^{s_{\pm}v}_{\tilde{1}'2}(u)$ are necessary to derive the exact
solution of the $D^{(2)}_3$ model. For simplicity, throughout this
paper, we denote $\tilde{1}'=1'$ for the representation $s_+$ and
$\tilde{1}'={\bar 1}'$ for the representation $s_-$. Here we list
some useful relations among the $R$-matrices:
\begin{eqnarray}
\hspace{-0.8truecm}{\rm unitarity}&:&R^{   s_{\pm}v}_{\tilde{1}'2}(u)R^{  vs_{\pm}}_{2\tilde{1}'}(-u)=\rho_{s}(u)=-4\sinh(u-3\eta)\sinh(u+3\eta),\nonumber\\[4pt]
\hspace{-0.8truecm}{\rm crossing \,\,unitarity}&:&R^{
s_{\pm}v}_{\tilde{1}'2}(u)^{t_{\tilde{1}'}}\bar{M}_{\tilde{1}'}R^{
 vs_{\pm}}_{2\tilde{1}'}(-u+8\eta)^{t_{\tilde{1}'}}\bar{M}_{\tilde{1}'}^{-1}=\rho_{s}(u-4\eta-i\pi), \nonumber\\[4pt]
 \hspace{-0.8truecm}{\rm periodicity}&:&R^{   s_{\pm}v}_{\tilde{1}'2}(u+i\pi)=-\bar{V}_2R^{
s_{\pm}v}_{\tilde{1}'2}(u)\bar{V}_2^{-1},\no %\label{Properties}
\end{eqnarray}where $\bar{M}$ is the diagonal matrix given by
\bea\bar{M}=diag[e^{4\eta},e^{2\eta},e^{-2\eta},e^{-4\eta}],\eea
and the Yang-Baxter relations \bea \hspace{-0.3truecm} R^{
s_{\pm}v}_{\tilde{1}'2}(u_1-u_2)R^{
s_{\pm}v}_{\tilde{1}'3}(u_1-u_3)R^{ vv}_{23}(u_2-u_3) =R^{
vv}_{23}(u_2-u_3)R^{ s_{\pm}v}_{\tilde{1}'3}(u_1-u_3)R^{
s_{\pm}v}_{\tilde{1}'2}(u_1-u_2).\label{QYB11-2}\eea

\subsection*{A.2 Fusion and the fused $R$-matrices}
By using the fusion technique \cite{Kul81,Kul86,Kar79,Kir86,Kir87,Mez92}, we systematically analyze the fusion structure of the $R$-matrices.
For this purpose,
we consider the fusion of $R^{s_+v}_{1'3}(u)$ and $R^{vv}_{23}(u)$
in the spaces ${\bf V}_{1'}$ and ${\bf V}_{2}$. This can be
realized is because  the fused $R$-matrix
$R^{s_{+}v}_{1'2}(u)$ (\ref{rsp}) has the degenerate point of
$u=3\eta+i\pi$. At which, $R^{s_{+}v}_{1'2}(u)$ becomes a $4\times
4$ matrix  \bea
R^{s_+v}_{1'2}(3\eta+i\pi)=P_{1'2}^{(+)}S_{1'2}^{(+)},\eea where
$P_{1'2}^{(+)}$ is a 4-dimensional projector \bea
P_{1'2}^{(+)}=\sum_{i=1}^{4}
|{\phi}^{(+)}_i\rangle\langle{\phi}^{(+)}_i|,\label{4-dim-1} \eea with the basis
vectors
 \bea
&&|{\phi}^{(+)}_1\rangle=e^{\frac{\eta}{2}}\frac{\sinh\eta\sqrt{\cosh
3\eta}}{\tilde{x}_1\sqrt{\cosh\eta\cosh
2\eta}}|13\rangle+e^{\frac{\eta}{2}}\frac{(e^{\eta}\cosh
2\eta-\sinh\eta)}{\tilde{x}_1\sqrt{\cosh 2\eta}}|14\rangle\no\\
&&\hspace{1.5truecm}+e^{{\eta}}\frac{\sqrt{\cosh
\eta}}{\tilde{x}_1}|22\rangle-e^{3{\eta}}\frac{\sqrt{\cosh
\eta}}{\tilde{x}_1}|31\rangle,\quad \tilde{x}_1=\sqrt{e^{3\eta}(2\cosh
2\eta+\cosh 4\eta)},\no\\
&&|{\phi}^{(+)}_2\rangle=\frac{\sqrt{\cosh
\eta}}{\tilde{x}_1}|15\rangle+e^{\frac{3\eta}{2}}\frac{\sqrt{\cosh
3\eta}}{\tilde{x}_1\sqrt{\cosh\eta\cosh 2\eta}}|23\rangle\no\\
&&\hspace{1.5truecm}+e^{\frac{3\eta}{2}}\frac{2\sinh^2\eta}{\tilde{x}_1\sqrt{\cosh
2\eta}}|24\rangle+e^{{3\eta}}\frac{\sqrt{\cosh
\eta}}{\tilde{x}_1}|41\rangle,\no\\
&&|{\phi}^{(+)}_3\rangle=\frac{\sqrt{\cosh
\eta}}{\tilde{x}_1}|16\rangle-e^{\frac{3\eta}{2}}\frac{\sqrt{\cosh
3\eta}}{\tilde{x}_1\sqrt{\cosh\eta\cosh 2\eta}}|33\rangle\no\\
&&\hspace{1.5truecm}-e^{\frac{3\eta}{2}}\frac{2\sinh^2\eta}{\tilde{x}_1\sqrt{\cosh
2\eta}}|34\rangle-e^{{3\eta}}\frac{\sqrt{\cosh
\eta}}{\tilde{x}_1}|42\rangle,\no\\
&&|{\phi}^{(+)}_4\rangle= \frac{\sqrt{\cosh
\eta}}{\tilde{x}_1}|26\rangle+e^{2\eta}\frac{\sqrt{\cosh
\eta}}{\tilde{x}_1}|35\rangle-e^{\frac{5\eta}{2}}\frac{\sinh\eta\sqrt{\cosh
3\eta}}{\tilde{x}_1\sqrt{\cosh\eta\cosh
2\eta}}|43\rangle\no\\
&&\hspace{1.5truecm}+e^{\frac{3\eta}{2}}\frac{(\cosh
2\eta+e^{\eta}\sinh\eta)}{\tilde{x}_1\sqrt{\cosh
2\eta}}|44\rangle.\eea $\tilde{x}_1=\sqrt{e^{3\eta}(2\cosh
2\eta+\cosh 4\eta)}$ and $S_{1'2}^{(+)}$ is a constant matrix
omitted here. Exchanging the two spaces ${\rm\bf V}_{1'}$ and
${\rm\bf V}_{2}$, we obtain the fused $R$-matrix
$R^{vs_{+}}_{21'}(u)$ . From it, we deduce another 4-dimensional
projector \bea P_{21'}^{(+)}=\sum_{i=1}^{4}
|{\varphi}^{(+)}_i\rangle\langle{\varphi}^{(+)}_i|,\quad
|{\varphi}^{(+)}_i\rangle=|{\phi}^{(+)}_i\rangle|_{\eta\rightarrow
-\eta,|kl\rangle\rightarrow |lk\rangle}. \eea Taking the fusion of
$R^{s_+v}_{1'3}(u)$ and $R^{vv}_{23}(u)$ by using the projectors
$P_{1'2}^{(+)}$ and $P^{ (+) }_{21'}$, we obtain \bea &&P^{ (+)
}_{1'2}R^{vv} _{23}(u)R^{s_+v} _{1'3}(u+3\eta+i\pi)P^{ (+)
}_{1'2}=\tilde{\rho}_0(u)R^{s_-v}
_{\langle 1'2\rangle 3}(u+\eta+i\pi), \label{sv-2-1} \\
&&P^{ (+) }_{21'}R^{vv} _{32}(u)R^{vs_+} _{31'}(u+3\eta+i\pi)P^{
(+) }_{21'}=\tilde{\rho}_0(u)R^{vs_-} _{3\langle
1'2\rangle}(u+\eta+i\pi), \label{sv-2}
 \eea
where $\tilde{\rho}_0(u)=4\sinh(u+2\eta)\sinh(u-4\eta)$, $\langle 1'2\rangle$ denotes the fused space ${\bf V}_{\langle 1'2\rangle}$,
and $R^{s_-v}_{\langle 1'2\rangle 3}(u)$ and $R^{vs_-}_{3\langle
1'2\rangle }(u)$ are the new fused $R$-matrices.
For simplicity, we define $\bar{1}'\equiv\langle 1'2\rangle$.
We shall note that although the dimension
of fused space ${\bf V}_{\bar{1}'}$ is 4, the ${\bf V}_{\bar{1}'}$ is not the original spinorial
representation space ${\bf V}_{1'}$, i.e.,  ${\bf V}_{1'}\neq {\bf V}_{\langle 1'2\rangle}$.
In fact, ${\bf V}_{\bar{1}'}$ is the space of
another spinorial representation $s_-$ of $D_3^{(2)}$ Lie algebra.
Thus the
fused $R$-matrix $R^{s_-v}_{\bar{1}'2}(u)$ is defined in the
tensor spaces of ${\bf V}_{\bar{1}'}\otimes {\bf V}_{2}$ and can
be expressed as a $24\times 24$ matrix.

The fused $R$-matrices (\ref{sv-2-1}) and (\ref{sv-2}) can also be used to make the further  fusion. At the point
of $u=3\eta+i\pi$, $R^{s_-v}_{\bar{1}'2}(u)$ reduces into a $4\times 4$ matrix \bea
R^{s_-v}_{\bar{1}'2}(3\eta+i\pi)=P_{\bar{1}'2}^{(-)}S_{\bar{1}'2}^{(-)},\eea
where $P_{\bar{1}'2}^{(-)}$ is a 4-dimensional projector \bea
P_{\bar{1}'2}^{(-)}=\sum_{i=1}^{4}
|{\phi}^{(-)}_i\rangle\langle{\phi}^{(-)}_i|, \label{4-dim-2}\eea with the basis
vectors \bea
&&|{\phi}^{(-)}_1\rangle=e^{\frac{\eta}{2}}\frac{\sqrt{\cosh\eta\cosh
3\eta}}{\tilde{x}_1\sqrt{\cosh
2\eta}}|13\rangle+e^{\frac{5\eta}{2}}\frac{\sinh
\eta}{\tilde{x}_1\sqrt{\cosh
2\eta}}|14\rangle\no\\
&&\hspace{1.5truecm}-e^{{\eta}}\frac{\sqrt{\cosh\eta}}{\tilde{x}_1}|22\rangle
-e^{3{\eta}}\frac{\sqrt{\cosh\eta}}{\tilde{x}_1}|31\rangle,\no\\
&&|{\phi}^{(-)}_2\rangle=\frac{\sqrt{\cosh\eta}}{\tilde{x}_1}|15\rangle
-e^{\frac{3\eta}{2}}\frac{\sqrt{\cosh
2\eta}}{\tilde{x}_1}|24\rangle+e^{{3\eta}}\frac{\sqrt{\cosh
\eta}}{\tilde{x}_1}|41\rangle,\no\\
&&|{\phi}^{(-)}_3\rangle=\frac{\sqrt{\cosh\eta}}{\tilde{x}_1}|16\rangle
-e^{\frac{3\eta}{2}}\frac{\sqrt{\cosh
2\eta}}{\tilde{x}_1}|34\rangle-e^{{3\eta}}\frac{\sqrt{\cosh
\eta}}{\tilde{x}_1}|42\rangle,\no\\
&&|{\phi}^{(-)}_4\rangle=\frac{\sqrt{\cosh\eta}}{\tilde{x}_1}|26\rangle
-e^{2{\eta}}\frac{\sqrt{\cosh\eta}}{\tilde{x}_1}|35\rangle
-e^{\frac{5\eta}{2}}\frac{\sqrt{\cosh\eta\cosh
3\eta}}{\tilde{x}_1\sqrt{\cosh
2\eta}}|43\rangle+\frac{e^{\frac{\eta}{2}}\sinh
\eta}{\tilde{x}_1\sqrt{\cosh 2\eta}}|44\rangle,\no\eea and $S_{\bar{1}'2}^{(-)}$ is a constant matrix omitted here. From
Eq.(\ref{sv-2}), we know that the fused $R$-matrix
$R^{vs_-}_{2\bar{1}'}(3\eta+i\pi)$ degenerates into the
4-dimensional projector \bea P_{2\bar{1}'}^{(-)}=\sum_{i=1}^{4}
|{\varphi}^{(-)}_i\rangle\langle{\varphi}^{(+)}_i|, \quad
|{\varphi}^{(-)}_i\rangle=|{\phi}^{(-)}_i\rangle|_{\eta\rightarrow
-\eta, |kl\rangle\rightarrow |lk\rangle}. \eea Taking the fusion
of $R^{s_-v}_{\bar{1}'3}(u)$  and $R^{vv}_{23}(u)$ by using the projectors $P_{\bar{1}'2}^{(-)}$ and
$P_{2\bar{1}'}^{(-)}$, we obtain \bea &&P^{ (-)
}_{\bar{1}'2}R^{vv} _{23}(u)R^{s_-v} _{\bar{1}'3}(u+3\eta+i\pi)P^{
(-) }_{\bar{1}'2}=\tilde{\rho}_0(u) \tilde{S}_{\langle
\bar{1}'2\rangle}R^{s_+v}
_{\langle \bar{1}'2\rangle 3}(u+\eta+i\pi)\tilde{S}_{\langle \bar{1}'2\rangle}^{-1}, \label{ai} \\
 &&P^{ (-) }_{2\bar{1}'}R^{vv} _{32}(u)R^{vs_-}
_{3\bar{1}'}(u+3\eta+i\pi)P^{ (-)
}_{2\bar{1}'}=\tilde{\rho}_0(u)\tilde{S}_{\langle
\bar{1}'2\rangle}R^{vs_+} _{3\langle \bar{1}'2\rangle
}(u+\eta+i\pi)\tilde{S}_{\langle \bar{1}'2\rangle}^{-1},
\label{sv-1111} \eea where $\langle {\bar 1}'2\rangle$ denotes the 4-dimensional fused space ${\bf V}_{\langle {\bar 1}'2\rangle}$ and  $\tilde{S}$ is a $4\times 4$ diagonal
matrix
\bea
\tilde{S}=diag[1,-1,1,-1].
\eea
We shall note that the space ${\bf V}_{\langle {\bar 1}'2\rangle}$
is indeed the original spin representation space ${\bf V}_{1'}$.
Then we have $1'=\langle {\bar 1}'2\rangle$. From Eqs.(\ref{ai})
and (\ref{sv-1111}), we know that the fusion of $R^{s_-
v}_{\bar{1}'3}(u)$ and $R^{v v}_{23}(u)$ gives the already
obtained fused $R$-matrix $R^{s_+ v}_{1'2}(u)$ (\ref{rsp}) and
(\ref{rsn}). Therefore, the fusion processes are closed.

At the point of $u=4\eta$, the $36\times 36$ dimensional $R$-matrix (\ref{Rv}) degenerates into
\bea
R^{vv}_{12}(4\eta)=P^{{  vv}(1) }_{12}S_{12}^{(1)}, \eea where
$P^{{  vv}(1) }_{12}$ is the
one-dimensional projector
\bea  P^{{ vv}(1)
}_{12}=|\psi_0\rangle\langle\psi_0|, \label{a1} \eea
with the basis vector
\bea &&|\psi_0\rangle=\frac{1}{x_0}
(e^{-3\eta}|16\rangle+e^{-\eta}|25\rangle-\sqrt{\frac{\cosh
3\eta}{\cosh\eta}}|34\rangle-
\sqrt{\frac{\cosh 3\eta}{\cosh\eta}}|43\rangle+e^{\eta}|52\rangle+e^{3\eta}|61\rangle),\no\\
&& x_0=\frac{\sqrt{\cosh\eta}}{\sqrt{2\cosh 3\eta(2\cosh
2\eta+\cosh 4\eta)}}, \eea and $S_{12}^{(1)}$ is a constant omitted here.
Exchanging two spaces ${\rm\bf V}_{1}$ and ${\rm\bf V}_{2}$, we
obtain \bea P^{{ vv}(1) }_{21}=
|\tilde{\psi}_0\rangle\langle\tilde{\psi}_0|, \quad
|\tilde{\psi}_0\rangle= |\psi_0\rangle|_{|kl\rangle\rightarrow |lk\rangle}.
\label{1a1} \eea The fusion of two vectorial $R$-matrices by using
the projectors $P^{{ vv}(1) }_{12}$ and $P^{{ vv}(1) }_{21}$ gives
\bea
&&P^{ {  vv}(1) }_{21}R^{vv} _{13}(u)R^{vv} _{23}(u+4\eta)P^{ {  vv}(1) }_{21}=a_1(u)e_1(u+4\eta)P^{{  vv} (1) }_{21},\label{hhhgg-1} \\
&&P^{ {  vv}(1) }_{12}R^{vv} _{31}(u)R^{vv}_{32}(u+4\eta)P^{ {
vv}(1) }_{12}=a_1(u)e_1(u+4\eta)P^{ {  vv}(1) }_{12},
\label{hhgg-1} \eea where $e_1(u)=4\sinh(u-2\eta)\,\sinh(u)$. We see
that after taking fusion, we obtain a one-dimensional vector.

At the point of $u=2\eta+i\pi$, the $R$-matrix (\ref{Rv}) reduces to a $16\times 16$ matrix \bea
R^{vv}_{12}(2\eta+i\pi)=P^{{ vv} (16) }_{12}S_{12}^{(16)},\eea
where $P^{{vv} (16) }_{12}$ is a 16-dimensional projector
\bea &&P^{{  vv}
(16) }_{12}=\sum_{i=1}^{16}
|{\phi}^{(16)}_i\rangle\langle{\phi}^{(16)}_i|, \label{16-dim}\eea
with the bases
\bea
&&|{{\phi}}^{(16)}_1\rangle=\frac{e^{-\eta}}{\sqrt{2\cosh
2\eta}}|12\rangle-\frac{e^{\eta}}{\sqrt{2\cosh
2\eta}} |21\rangle,\no\\
&&|{{\phi}}^{(16)}_2\rangle=\frac{e^{-\eta}\sqrt{\cosh
2\eta}}{x_1}|13\rangle-\frac{e^{2\eta}\sinh\eta}{x_1\sqrt{\cosh
2\eta}}|31\rangle+\frac{\sqrt{\cosh \eta\cosh
3\eta}}{x_1\sqrt{\cosh
2\eta}}|41\rangle,\no\\&&|{{\phi}}^{(16)}_3\rangle=\frac{\sinh\eta\sqrt{\cosh
\eta\cosh 3\eta}}{x_1{\cosh
2\eta}}|13\rangle+\frac{\cosh\eta\sqrt{\cosh \eta\cosh
3\eta}}{x_1{\cosh 2\eta}}|31\rangle\no\\&&
\hspace{15mm}+\frac{x_1}{2{\cosh
2\eta}}|14\rangle+\frac{\sinh\eta\sinh 3\eta}{x_1{\cosh
2\eta}}|41\rangle,\no\\&&
|{{\phi}}^{(16)}_4\rangle=\frac{e^{-\eta}}{\sqrt{2\cosh
2\eta}}|15\rangle-\frac{e^{\eta}}{\sqrt{2\cosh 2\eta}}
|51\rangle,\no\\&&
|{{\phi}}^{(16)}_5\rangle=\frac{e^{-\frac{\eta}{2}}\cosh\eta}{x_2\sqrt{\cosh
2\eta}} |16\rangle+\frac{e^{-\frac{\eta}{2}}\cosh
2\eta}{x_2\sqrt{\cosh 2\eta}}
|44\rangle+\frac{e^{\frac{3\eta}{2}}\cosh\eta}{x_2\sqrt{\cosh
2\eta}} |52\rangle,\no\\
 &&|{{\phi}}^{(16)}_6\rangle=\frac{e^{-\eta}\sqrt{\cosh
2\eta}}{x_1}|23\rangle-\frac{e^{2\eta}\sinh\eta}{x_1\sqrt{\cosh
2\eta}}|32\rangle+\frac{\sqrt{\cosh \eta\cosh
3\eta}}{x_1\sqrt{\cosh 2\eta}}|42\rangle,\no\\&&
|{{\phi}}^{(16)}_7\rangle=\frac{\sinh\eta\sqrt{\cosh \eta\cosh
3\eta}}{x_1{\cosh 2\eta}}|23\rangle+\frac{\cosh\eta\sqrt{\cosh
\eta\cosh 3\eta}}{x_1{\cosh
2\eta}}|32\rangle\no\\&&\hspace{15mm}+\frac{x_1}{2{\cosh
2\eta}}|24\rangle+\frac{\sinh\eta\sinh 3\eta}{x_1{\cosh
2\eta}}|42\rangle,\no\\&&
|{{\phi}}^{(16)}_8\rangle=-\frac{e^{-\frac{\eta}{2}}\cosh
2\eta}{2x_2x_3\sqrt{\cosh 2\eta}}
|16\rangle+\frac{e^{-\frac{3\eta}{2}}x_2}{2x_3\sqrt{\cosh 2\eta}}
|25\rangle+\frac{e^{\frac{3\eta}{2}}\sinh
4\eta}{4x_2x_3\sinh\eta\sqrt{\cosh 2\eta}} |44\rangle,\no\\
&&\hspace{15mm}-\frac{e^{\frac{3\eta}{2}}\cosh
2\eta}{2x_2x_3\sqrt{\cosh 2\eta}}
|52\rangle+\frac{e^{\frac{\eta}{2}}x_2}{2x_3\sqrt{\cosh
2\eta}} |61\rangle, \no\\
&&|{{\phi}}^{(16)}_9\rangle=\frac{e^{-\eta}}{\sqrt{2\cosh
2\eta}}|26\rangle-\frac{e^{\eta}}{\sqrt{2\cosh
2\eta}} |62\rangle,\no\\
&&|{{\phi}}^{(16)}_{10}\rangle=\frac{(e^{\eta}\cosh^2
2\eta+\sinh\eta)\sqrt{\cosh\eta}}{2x_3x_4{\cosh^{\frac32}
2\eta}}|16\rangle+\frac{e^{2\eta}(e^{\eta}\cosh^2
2\eta+\sinh\eta)\sqrt{\cosh\eta}}{2x_3x_4{\cosh^{\frac32}
2\eta}}|25\rangle\no\\
&&\hspace{15mm}+\frac{e^{-3\eta}(\cosh^2
2\eta-e^{\eta}\sinh\eta)\sqrt{\cosh\eta}}{2x_3x_4{\cosh^{\frac32}
2\eta}}|52\rangle\no\\
&&\hspace{15mm}+\frac{e^{-\eta}(\cosh^2
2\eta-e^{\eta}\sinh\eta)\sqrt{\cosh\eta}}{2x_3x_4{\cosh^{\frac32}
2\eta}}|61\rangle\no\\
&&\hspace{15mm}+\frac{{x_3\cosh 3\eta}}{x_4\sqrt{\cosh\eta\cosh
2\eta}}|33\rangle-\frac{{\cosh^{\frac32} \eta}}{x_3x_4\sqrt{\cosh
2\eta}}|44\rangle,\no\\
&&|{{\phi}}^{(16)}_{11}\rangle=-\frac{e^{-\eta}(2e^{\eta}+\cosh
5\eta-\sinh 3\eta)\sqrt{\cosh\eta\cosh
3\eta}}{2x_4x_5{\cosh^{\frac32}
2\eta}}|16\rangle\no\\
&&\hspace{15mm}-\frac{e^{\eta}(2e^{\eta}+\cosh 5\eta-\sinh
3\eta)\sqrt{\cosh\eta\cosh 3\eta}}{2x_4x_5{\cosh^{\frac32}
2\eta}}|25\rangle\no\\
&&\hspace{15mm}+\frac{e^{-\eta}(2e^{-\eta}+\cosh 5\eta+\sinh
3\eta)\sqrt{\cosh\eta\cosh 3\eta}}{2x_4x_5{\cosh^{\frac32}
2\eta}}|52\rangle\no\\
&&\hspace{15mm}+\frac{e^{\eta}(2e^{-\eta}+\cosh 5\eta+\sinh
3\eta)\sqrt{\cosh\eta\cosh 3\eta}}{2x_4x_5{\cosh^{\frac32}
2\eta}}|61\rangle\no\\
&&\hspace{15mm}+\frac{2\sinh \eta\sqrt{\cosh\eta\cosh 2\eta\cosh
3\eta}}{x_4x_5}|33\rangle+\frac{x_4\sinh \eta}{x_5\sqrt{\cosh
2\eta}}|34\rangle\no\\
&&\hspace{15mm}+\frac{x_4\sinh \eta}{x_5\sqrt{\cosh
2\eta}}|43\rangle-\frac{2\cosh\eta\sinh 2\eta\sqrt{\cosh\eta\cosh
3\eta}}{x_4x_5\sqrt{\cosh
2\eta}}|44\rangle,\no\\
&&|{{\phi}}^{(16)}_{12}\rangle=\frac{e^{-\eta}\sqrt{\cosh
2\eta}}{x_1}|35\rangle-\frac{e^{2\eta}\sinh\eta}{x_1\sqrt{\cosh
2\eta}}|53\rangle+\frac{\sqrt{\cosh \eta\cosh
3\eta}}{x_1\sqrt{\cosh 2\eta}}|54\rangle,\no\\&&
|{{\phi}}^{(16)}_{13}\rangle=\frac{e^{-\eta}\sqrt{\cosh
2\eta}}{x_1}|36\rangle-\frac{e^{2\eta}\sinh\eta}{x_1\sqrt{\cosh
2\eta}}|63\rangle+\frac{\sqrt{\cosh
\eta\cosh 3\eta}}{x_1\sqrt{\cosh 2\eta}}|64\rangle,\no\\
&& |{{\phi}}^{(16)}_{14}\rangle=\frac{\sinh\eta\sqrt{\cosh
\eta\cosh 3\eta}}{x_1{\cosh
2\eta}}|35\rangle+\frac{\cosh\eta\sqrt{\cosh
\eta\cosh 3\eta}}{x_1{\cosh 2\eta}}|53\rangle\no\\
&&\hspace{15mm}+\frac{x_1}{2{\cosh
2\eta}}|45\rangle+\frac{\sinh\eta\sinh
3\eta}{x_1{\cosh 2\eta}}|54\rangle,\no\\
&& |{{\phi}}^{(16)}_{15}\rangle=\frac{\sinh\eta\sqrt{\cosh
\eta\cosh 3\eta}}{x_1{\cosh
2\eta}}|36\rangle+\frac{\cosh\eta\sqrt{\cosh
\eta\cosh 3\eta}}{x_1{\cosh 2\eta}}|63\rangle\no\\
&&\hspace{15mm}+\frac{x_1}{2{\cosh
2\eta}}|46\rangle+\frac{\sinh\eta\sinh
3\eta}{x_1{\cosh 2\eta}}|64\rangle,\no\\
&&|{{\phi}}^{(16)}_{16}\rangle=\frac{e^{-\eta}}{\sqrt{2\cosh
2\eta}}|56\rangle-\frac{e^{\eta}}{\sqrt{2\cosh 2\eta}}
|65\rangle,\no\\
&&x_1=\sqrt{1+\cosh 4\eta-\sinh 2\eta}, \quad  x_2=\sqrt{e^{\eta}+\cosh\eta\cosh 2\eta},\no\\
&& x_3=\sqrt{1+2\cosh\eta\cosh 3\eta}, \qquad  x_4=\sqrt{\cosh 3\eta+2\cosh\eta\cosh 2\eta\cosh 4\eta},\no\\
&& x_5=\sqrt{3\cosh 3\eta+2\sinh\eta\sinh 2\eta\cosh 4\eta},
\eea
and $S_{12}^{(16)}$ is a constant matrix omitted here.
Exchanging the spaces ${\rm\bf V}_{1}$ and ${\rm\bf V}_{2}$, we obtain
 \bea P^{{  vv}
(16) }_{21}=\sum_{i=1}^{16}
|{\varphi}^{(16)}_i\rangle\langle{\varphi}^{(16)}_i|, \quad
|{{\varphi}}^{(16)}_i\rangle=|{{\phi}}^{(16)}_i\rangle|_{\eta\rightarrow-\eta,\, |kl\rangle\rightarrow|lk\rangle}. \eea
Taking the fusion of two vectorial $R$-matrices by using the
16-dimensional projectors $P^{{ vv} (16) }_{12}$ and $P^{{ vv} (16) }_{21}$, we obtain
\bea &&P^{{  vv} (16) }_{12}R^{vv}
_{23}(u)R^{vv}
_{13}(u+2\eta+i\pi)P^{{  vv} (16) }_{12}\no\\
&&\qquad\qquad=\tilde{\rho}_0(u) S_{1'\bar{2}'}R^{ s_+v}_{1'3}(u+\eta+i\pi)R^{
s_-v}_{\bar{2}'3}(u+\eta+i\pi)S_{1'\bar{2}'}^{-1}, \label{uf-12} \\
  &&P^{{  vv} (16) }_{21}R^{vv} _{32}(u)R^{vv}
_{32}(u+2\eta+i\pi)P^{{  vv} (16) }_{12}\no\\
&&\qquad\qquad=\tilde{\rho}_0(u) \bar{S}_{1'\bar{2}'}R^{
vs_+}_{31'}(u+\eta+i\pi)R^{
vs_-}_{3\bar{2}'}(u+\eta+i\pi)\bar{S}_{1'\bar{2}'}^{-1}.
\label{fu-12}\eea Here, two 16-dimensional
spaces ${\rm\bf V}_1$ and ${\rm\bf V}_2$ are fused into a 16-dimensional fused space ${\rm\bf V}_{\langle 12 \rangle}$. We
find that the fused space ${\rm\bf V}_{\langle 12
\rangle}$ can be divided into two
4-dimensional spaces ${\rm\bf V}_{1'}$ and ${\rm\bf
V}_{\bar{2}'}$, where ${\rm\bf V}_{1'}$ is the space of spinorial representation $s_+$ and ${\rm\bf
V}_{\bar{2}'}$ is the space of spinorial representation $s_-$. The $S_{1'\bar{2}'}$ is a
$16\times 16$ matrix defined in the tensor space ${\rm\bf V}_{1'}\otimes {\rm\bf V}_{\bar{2}'}$ \bea
 &&\hspace{-2.0cm}S_{1'\bar{2}'}=\left(\begin{array}{cccc|cccc|cccc|cccc}
    1&&& &&&& &&&& &&&& \\
    &s_1&& &s_2&&& &&&& &&&& \\
    &s_3&& &s_4&&& &&&& &&&& \\
    &&& &&-1&& &&&& &&&& \\
   \hline &&&s_5 &&&s_6& &&s_7&& &s_8&&& \\
    &&-s_1& &&&& &s_2&&& &&&& \\
    &&-s_3& &&&& &s_4&&& &&&& \\
    &&&s_9 &&&s_{10}& &&&& &s_{11}&&& \\
   \hline &&& &&&& &&&-1& &&&& \\
    &&&s_{12} &&&& &&&& &s_{13}&&& \\
    &&&s_{14} &&&& &&&& &s_{15}&&& \\
    &&& &&&&s_1  &&&& &&s_2&& \\
   \hline &&& &&&& &&&& -s_1&&&s_2& \\
    &&& &&&&s_3 &&&& &&s_4&& \\
    &&& &&&& &&&&-s_3 &&&s_4& \\
    &&& &&&& &&&& &&&&-1 \\
           \end{array}\right),\label{Tra1}
\eea
where the nonzero matrix elements are
 \bea&&\hspace{-0.5cm}
s_1=-\frac{\cosh\eta}{x_1}\sqrt{\frac{2\cosh 3\eta}{e^{\eta}\cosh
2\eta}},\ s_2=\frac{\sinh\eta}{x_1}\sqrt{\frac{2e^{\eta}\cosh
3\eta}{\cosh
2\eta}},\  s_3=\frac{\sinh\eta}{x_1}\sqrt{\frac{2e^{3\eta}}{\cosh
\eta}}, \no\\[6pt]
&& \hspace{-0.5cm} s_4=-\frac{\cosh
2\eta}{x_1}\sqrt{\frac{2}{e^{\eta}\cosh \eta}},\
 s_5=-\frac{\sinh\eta(e^{-\eta}+\cosh 3\eta-\sinh\eta)}{x_2\sinh
4\eta}\sqrt{2e^{5\eta}},\no \\&& \hspace{-0.5cm} s_6=\frac{\cosh
2\eta}{x_2\cosh\eta}\sqrt{\frac{e^{\eta}}{2}},\ s_7=\frac{x_2}{\cosh\eta}\sqrt{\frac{1}{2e^{\eta}}},\
s_8=\frac{\sinh\eta(e^{\eta}+\cosh 3\eta+\sinh\eta)}{x_2\sinh
4\eta}\sqrt{\frac{2}{e^{7\eta}}},\no\\[6pt]
&&\hspace{-0.5cm} s_9=\frac{e^{2\eta}\cosh\eta}{x_3}s_5,\
s_{10}=\frac{x_2}{x_3}\sqrt{2e^{\eta}},\
s_{11}=\frac{e^{2\eta}\cosh\eta}{x_3}s_{8},\no\\[6pt]
&& \hspace{-0.5cm}s_{12}=-\frac{\cosh 3\eta+\cosh 2\eta(e^{\eta}+\cosh
5\eta-\sinh 3\eta)}{x_3x_4\cosh
2\eta}\sqrt{\frac{\cosh\eta}{2}},\no\\[6pt]
&&\hspace{-0.5cm}s_{13}=\frac{\cosh 3\eta+\cosh 2\eta(e^{-\eta}+\cosh 5\eta+\sinh
3\eta)}{x_3x_4\cosh
2\eta}\sqrt{\frac{\cosh\eta}{2}},\no\\[6pt]
&&\hspace{-0.5cm}s_{14}=-e^{2\eta}\frac{\cosh 3\eta\sinh 6\eta+2\sinh
2\eta\sinh^2\eta(2\cosh{2\eta}-1)x_4^2}{x_4x_5\cosh
2\eta\sinh 4\eta}\sqrt{\frac{2\cosh\eta}{\cosh 3\eta}},\no\\[6pt]
&&\hspace{-0.5cm}s_{15}=-\frac{x_5}{2e^{2\eta}x_4\cosh
2\eta}\sqrt{\frac{2\cosh3\eta}{\cosh\eta}}.\eea The matrix
$\bar{S}_{1'\bar{2}'}$ can be obtained from the ${S}_{1'\bar{2}'}$
by using the mapping
\bea\bar{S}_{1'\bar{2}'}=S_{1'\bar{2}'}|_{\eta\rightarrow
-\eta}.\label{Tra2}
\eea

%%%%%%%%%%%%%%%%%%%%%%%%%%%%%%%%%%%%%%%%%%%%%%%%%%%%%%%%%%%%%%%%
%                                                             %
%  Appendix B                                                 %
%                                                             %
%                                                             %
%                                                             %
%%%%%%%%%%%%%%%%%%%%%%%%%%%%%%%%%%%%%%%%%%%%%%%%%%%%%%%%%%%%%%%

\section*{Appendix B: Fusion of the reflection matrices }
\setcounter{equation}{0}
\renewcommand{\theequation}{B.\arabic{equation}}
In this appendix, we shall give the related fusion of the reflection matrices \cite{Mez92-1, Zho96} associated with those of the $R$-matrices in Appendix A.
By using the one-dimensional projector $P_{12}^{ vv(1)}$, we obtain the fusion relation of reflection matrix $K^{ v}(u)$ and that of $\bar K^{ v}(u)$ as
\bea && P_{21}^{ vv(1)}K_{1}^{ v}(u)R_{21}^{ vv}(2u+4\eta)K_{1}^{ v}(u+4\eta)P_{12}^{ vv(1)}\no\\[4pt]
&&\qquad  =8\sinh(u+4\eta)\sinh(2u+6\eta)\cosh(u)\cosh(u+\eta)\cosh(u-\eta) P_{12}^{ vv(1)}, \label{8005-1}\\[4pt]
&& P_{12}^{ vv(1)}\bar{K}_{2}^{ v}(u+4\eta)M_1R_{12}^{ vv}(-2u+4\eta)M_1^{-1}
\bar{K}_{1}^{ v}(u)P_{21}^{\rm vv(1)}\no\\[4pt]
&& \qquad
=8\sinh(u-4\eta)\sinh(2u-6\eta)\cosh(u)\cosh(u+\eta)\cosh(u-\eta)P_{21}^{
    vv(1)}. \label{0805-1} \eea
We see that the fused results are the one-dimensional vectors.
Taking the fusion of reflection matrices by using the 16-dimensional projector $P_{12}^{ vv(16)}$, we have
\bea && P_{12}^{ vv(16)}K^{
    v}_2(u)R^{ vv}_{12}(2u+2\eta+i\pi)K^{ v}_1(u+2\eta+i\pi)P_{21}^{ vv(16)}\no\\[4pt]
    &&\quad =-16\cosh^2 2\eta\cosh(u+\eta)\sinh(u+2\eta)\no\\[4pt]
    &&\qquad \times S_{1'\bar{2}'}K^{
    s_+}_{1'}(u+\eta+i\pi)R^{ s_-s_+}_{\bar{2}'1'}(2u+2\eta+2i\pi)K^{
    s_-}_{\bar{2}'}(u+\eta+i\pi)\bar{S}_{1'\bar{2}'}^{-1}, \label{8005-2} \\[4pt]
&&P^{ vv(16)}_{21}\bar{K}^{
    v}_1(u+2\eta+i\pi)M_1^{-1}R^{ vv}_{21}(-2u+6\eta-i\pi)M_1\bar{K}^{v}_2(u)P^{ vv(16)}_{12}\no\\[4pt]
    &&\quad  =16\cosh^2 2\eta\cosh(u-3\eta)\sinh(u-4\eta)\no\\[4pt]
    &&\qquad \times\bar{S}_{1'\bar{2}'}\bar{K}^{
    s_-}_{\bar{2}'}(u+\eta+i\pi)\bar{M}_{\bar{2}'}^{-1}R^{ s_+s_-}_{1'\bar{2}'}(-2u+6\eta)\bar{M}_{\bar{2}'}\bar{K}^{
    s_+}_{1'}(u+\eta+i\pi)S_{1'\bar{2}'}^{-1}.  \label{0805-2} \eea
From Eqs.(\ref{8005-2}) and (\ref{0805-2}), we know that the
36-dimensional auxiliary spaces ${\rm\bf V}_1$ and ${\rm\bf V}_2$
are  projected into a 16-dimensional fused space ${\rm\bf
V}_{\langle 12\rangle}$, which can be divided into two 4-dimensional spaces ${\rm\bf V}_{1'}$ and ${\rm\bf
V}_{\bar{2}'}$, i.e., ${\rm\bf V}_{\langle 12 \rangle}={\rm\bf
V}_{1'}\otimes {\rm\bf V}_{\bar{2}'}$. As mentioned before,
${\rm\bf V}_{1'}$ is the space of spinorial representation $s_+$ and ${\rm\bf V}_{\bar{2}'}$ is
the space of spinorial representation
$s_-$. All the matrices in the right hand side of
Eqs.(\ref{8005-2}) and (\ref{0805-2}) are defined in the tensor
space ${\rm\bf V}_{1'}\otimes {\rm\bf V}_{\bar{2}'}$. The matrices
$S_{1'\bar{2}'}$ and $\bar S_{1'\bar{2}'}$ are given by
(\ref{Tra1}) and (\ref{Tra2}), respectively. The
$R^{s_+s_-}_{1'\bar{2}'}(u)$ is another $16\times 16$ spinorial $R$-matrix of the $D^{(2)}_3$
model \bea
 &&\hspace{-1.0cm}R^{s_+s_-}_{1'\bar{2}'}(u)=\no\\&&\hspace{-1.0cm}
 \left(\begin{array}{cccc|cccc|cccc|cccc}
    r_1&&& &&&& &&&& &&&& \\
    &r_2&& &r_3&&& &&&& &&&& \\
    &&r_2& &&&& &-r_3&&& &&&& \\
    &&&r_5 &&&-r_7& &&r_6&& &r_{10}&&& \\
   \hline &r_{4}&& &r_{2}&&& &&&& &&&& \\
    &&& &&r_{1}&& &&&& &&&& \\
    &&&-r_{9} &&&r_{5}& &&r_{12}&& &-r_{6}&&& \\
    &&& &&&&r_{2} &&&& &&r_{3}&& \\
   \hline &&-r_{4}& &&&& &r_{2}&&& &&&& \\
    &&&-r_{8} &&&r_{13}& &&r_{5}&& &r_{7}&&& \\
    &&& &&&& &&&r_{1}& &&&& \\
    &&& &&&&  &&&&r_{2} &&&-r_{3}& \\
   \hline &&&r_{11} &&&r_{8}& &&r_{9}&& &r_{5}&&& \\
    &&& &&&&r_{4} &&&& &&r_{2}&& \\
    &&& &&&& &&&&-r_{4} &&&r_{2}& \\
    &&& &&&& &&&& &&&&r_{1} \\
           \end{array}\right),\label{Rsspm}
\eea where the non-zero matrix elements are
 \bea
 &&r_{1}=\sinh(\frac{u}{2}-2\eta)\cosh(\frac u2-\eta),\quad
r_{2}=\sinh(\frac{u}{2}-2\eta)\cosh \frac u2,\no\\[6pt]
 &&r_{3}=e^{-\frac u2}\sinh\eta\sinh(\frac{u}{2}-2\eta),\quad r_{4}=-e^{\frac u2}\sinh\eta\sinh(\frac{u}{2}-2\eta),\no\\[6pt]
 && r_{5}=\sinh(\frac{u}{2}-\eta)\cosh \frac u2,\quad r_{6}=-e^{-\frac u2-\eta}\sinh\eta\cosh \frac{u}{2},\no\\[6pt]
&&r_{7}=e^{-\frac u2+\eta}\sinh\eta\cosh \frac{u}{2},\quad
r_{8}=e^{u} r_7, \quad r_{9}=e^{u-2\eta} r_7,\no\\[6pt]
 &&r_{10}=-e^{-u}\sinh\eta\cosh (2\eta),\quad r_{11}=e^{2u}r_{10}, \quad
r_{12}=e^{-\frac u2}\sinh\eta
 [\cosh \frac u2 -\sinh(\frac u2-2\eta)],\no\\ &&r_{13}=e^{\frac u2}\sinh\eta
 [\cosh \frac u2 +\sinh(\frac u2-2\eta)].
 \eea
The $K^{s_\pm}(u)$ are the spinorial reflection matrices defined in the space ${\rm\bf V}_{{\tilde k}'}$
\bea
K^{s_\pm}(u)=\left(\begin{array}{cccc}e^{-u}&0&0&c\sinh u\\
    0&\frac{\cosh(u-2\eta)}{\cosh 2\eta}&0&0\\
    0&0&\frac{\cosh(u-2\eta)}{\cosh 2\eta}&0\\
    -\frac{\sinh u}{c\cosh^2 2\eta}&0&0&e^{u}\end{array}\right).\label{K-matrix-1}
\eea
From the above equation, we see that two spinorial reflection matrices $K^{s_\pm}(u)$ are the same, $K^{s_+}(u)=K^{s_-}(u)$,
although they are defined in different spaces. The $\bar{K}^{s_{\pm}}(u)$ are the dual
reflection matrices with the definition \bea
\bar{K}^{s_{\pm}}(u)=\bar{M}K^{s_{\pm}}(-u+4\eta+i\pi)\left|_{c\rightarrow
   c'}\right.. \label{K-matrix-spinor-1}\eea

The spinorial $R$-matrix $R^{ s_+s_-}_{1'\bar{2}'}(u)$ has the following properties:
\bea &&\hspace{-1cm} {\rm unitarity}:R^{
s_+s_-}_{1'\bar{2}'}(u)R^{
s_-s_+}_{\bar{2}'1'}(-u)=\rho_{ss}(u),\\
&&\qquad \;\;\rho_{ss}(u)=-\sinh(\frac u2+2\eta)\sinh(\frac
u2-2\eta)\cosh(\frac u2+\eta)\cosh(\frac u2-\eta),\no \\
&&\hspace{-1cm} {\rm crossing\,\,unitarity}:R^{
s_+s_-}_{1'\bar{2}'}(u)^{t_{1'}}\bar{M}_{1'}R^{
 s_-s_+}_{\bar{2}'1'}(-u+8\eta)^{t_{1'}}\bar{M}_{1'}^{-1}=\rho_{ss}(u-4\eta-i\pi),\eea
and satisfies the Yang-Baxter equation \bea &&R^{
s_+s_-}_{1'\bar{2}'}(u_1-u_2)R^{ s_+v}_{1'3}(u_1-u_3)R^{
s_-v}_{\bar{2}'3}(u_2-u_3)\no \\
&&\qquad\qquad =R^{ s_-v}_{\bar{2}'3}(u_2-u_3)R^{ s_+v}_{1'3}(u_1-u_3)R^{
s_+s_-}_{1'\bar{2}'}(u_1-u_2). \nonumber
\eea The matrices $K^{s_{\pm}}(u)$
satisfy the reflection equation \bea && R^{
s_{\pm}v}_{\tilde{1}'2}(u-v){K^{  s_{\pm}}_{  \tilde{1}'}}(u)
R^{   vs_{\pm}}_{2\tilde{1}'}(u+v) {K^{   v}_{2}}(v) \no \\
&&\qquad\qquad=
 {K^{   v}_{2}}(v)R^{   s_{\pm}v}_{\tilde{1}'2}(u+v){K^{   s_{\pm}}_{\tilde{1}'}}(u)
 R^{   vs_{\pm}}_{2\tilde{1}'}(u-v).
 \label{1r2}
 \eea
The matrices $\bar K^{s_{\pm}}(u)$ satisfy the dual reflection
equation \bea &&R^{s_{\pm}v}_{\tilde{1}'2}(-u+v){\bar{K}^{
s_{\pm}}_{\tilde{1}'}}(u)\bar{M}_{\tilde{1}'}^{-1}R^{
vs_{\pm}}_{2\tilde{1}'}
 (-u-v+8\eta+2i\pi)\bar{M}_{\tilde{1}'}{\bar{K}^{   v}_{2}}(v)\nonumber\\
&&\qquad\qquad={\bar{K}^{ v}_{2}}(v)\bar{M}_{\tilde{1}'}R^{
s_{\pm}v}_{\tilde{1}'2}(-u-v+8\eta+2i\pi)\bar{M}_{\tilde{1}'}^{-1}
{{\bar{K}}^{s_{\pm}}_{\tilde{1}'}}(u)R^{
vs_{\pm}}_{2\tilde{1}'}(-u+v).
 \label{1dr3}
 \eea

Taking the fusion of reflection matrices $K^{ v}(u)$ and
$K^{s_\pm}(u)$ by using the 4-dimensional projector
$P_{\tilde{1}'2}^{(\pm)}$, we obtain \bea &&
P_{1'2}^{(+)}K_2^{v}(u)R_{1'2}^{ s_{+}v}(2u+3\eta+i\pi)K_{1'}^{
s_{+}}(u+3\eta+i\pi) P_{21'}^{({+})}
\no\\&&\quad\quad=-4\cosh(u)\cosh(u+\eta)\sinh(u+3\eta)
K_{\langle 1'2\rangle }^{ s_{-}}(u+\eta+i\pi),\label{10805-31}\\
&& P_{\bar{1}'2}^{(-)}K_2^{v}(u)R_{\bar{1}'2}^{
s_{-}v}(2u+3\eta+i\pi)K_{\bar{1}'}^{ s_{-}}(u+3\eta+i\pi)
P_{2\bar{1}'}^{({-})}
\no\\&&\quad\quad=-4\cosh(u)\cosh(u+\eta)\sinh(u+3\eta)\tilde{S}_{\langle
\bar{1}'2\rangle} K_{\langle \bar{1}'2\rangle }^{
s_{+}}(u+\eta+i\pi)\tilde{S}_{\langle
\bar{1}'2\rangle}^{-1}.\label{20805-31} \eea From
Eqs.(\ref{10805-31}) and (\ref{20805-31}), we see that the
reflection matrix $K^{s_\mp}(u)$ can be obtained from $K^{ v}(u)$
and $K^{s_\pm}(u)$. Similarly, the fusion of dual reflection
matrices $\bar K^{ v}(u)$ and $\bar K^{s_\pm}(u)$ by using the
4-dimensional projector $P_{2\tilde{1}'}^{(\pm)}$ gives \bea &&
P_{21'}^{(+)}\bar{K}_{1'}^{
s_{+}}(u+3\eta+i\pi)\bar{M}_{1'}^{-1}R_{21'}^{
vs_{+}}(-2u+5\eta+i\pi)\bar{M}_{1'}\bar{K}_2^{v}(u)P_{1'2}^{(+)}
\no\\&&\qquad\qquad=
4\cosh(u+\eta)\cosh(u-3\eta)\sinh(u-4\eta)\bar{K}_{\langle
1'2\rangle }^{ s_{-}}(u+\eta+i\pi), \label{0805-3} \\
&& P_{2\bar{1}'}^{(-)}\bar{K}_{\bar{1}'}^{
s_{-}}(u+3\eta+i\pi)\bar{M}_{\bar{1}'}^{-1}R_{2\bar{1}'}^{
vs_{-}}(-2u+5\eta+i\pi)\bar{M}_{\bar{1}'}\bar{K}_2^{v}(u)P_{\bar{1}'2}^{(-)}
\no\\&&\quad\quad=
4\cosh(u+\eta)\cosh(u-3\eta)\sinh(u-4\eta)\tilde{S}_{\langle
\bar{1}'2\rangle}\bar{K}_{\langle \bar{1}'2\rangle }^{
s_{+}}(u+\eta+i\pi)\tilde{S}_{\langle \bar{1}'2\rangle}^{-1}.
\label{0805-31} \eea From Eqs.(\ref{0805-3}) and (\ref{0805-31}),
we see that the dual reflection matrix $\bar K^{s_\mp}(u)$ can be
obtained from $\bar K^{ v}(u)$ and $\bar K^{s_\pm}(u)$, which means that the fusion process now is closed.

%%%%%%%%%%%%%%%%%%%%%%%%%%%%%%%%%%%%%%%%%%%%%%%%%%%%%%%%%%%%%%%
%                                                             %
%  References                                                 %
%                                                             %
%%%%%%%%%%%%%%%%%%%%%%%%%%%%%%%%%%%%%%%%%%%%%%%%%%%%%%%%%%%%%%%

\end{document}